\documentclass[preprint,12pt]{elsarticle}
\usepackage{comment}
\usepackage{color}

\usepackage{amsthm, amsmath, amssymb, bm}

\usepackage{booktabs}
\usepackage{multirow}
\usepackage{url}
\usepackage{tabularx}
\usepackage{array}
\usepackage{makecell}

\newcolumntype{Y}{>{\raggedright\arraybackslash}X}

\usepackage{algorithm,algpseudocode}

\journal{Journal of Parallel and Distributed Computing}

\begin{document}

\begin{frontmatter}

\title{MERBIT: A GPU-Based SpMV Method for Iterative Workloads}
\tnotetext[funding]{This work was supported by the National Key Research and Development Program of China under Grant No. 2020YFA0712500, and the Guangdong Basic and Applied Basic Research Foundation under Grant Nos. 2022A1515010900 and 2024A1515012286.}

\author[inst1]{Qi Zhang}
\author[inst1]{Zhengan Yao}
\author[inst1]{Zhenglu Jiang}
\author[inst2]{Zan-Bo Zhang\corref{cor1}}
\ead{zanbozhang@gdufe.edu.cn}

\cortext[cor1]{Corresponding author}

\affiliation[inst1]{
	organization={Department of Mathematics},
	addressline={Sun Yat-sen University},
	city={Guangzhou},
	country={China}
}

\affiliation[inst2]{
	organization={School of Statistics and Data Science},
	addressline={Guangdong University of Finance and Economics},
	city={Guangzhou},
	country={China}
}

\begin{abstract}
Sparse Matrix–Vector Multiplication (SpMV) is the cornerstone in many iterative workloads, including large-scale graph analytics and sparse iterative solvers. Accelerating SpMV on real-world graphs remains challenging due to highly irregular sparsity patterns. In this paper, we propose MERBIT, a GPU SpMV method designed for repeated SpMV on irregular, graph-like sparse matrices, with PageRank as a representative motivating workload. MERBIT combines two key ideas from existing GPU SpMV methods. At the global level, it uses merge-path partitioning to balance work over nonzeros and row boundaries. At the local level, it encodes each merge-path segment using a compact bit-field descriptor. MERBIT improves workload balance and promotes coalesced memory access for both matrix loading and output writes; moreover, three optimization strategies are incorporated to further enhance performance. Experiments on 50 large irregular datasets demonstrate that MERBIT outperforms competitive baselines, including cuSPARSE, Ginkgo, and academic approaches, achieving average speedups of 1.27× and 1.25× over cuSPARSE COO in single and double precision, respectively.
\end{abstract}


\end{frontmatter}

\section{Introduction}
Sparse Matrix–Vector Multiplication (SpMV) is the cornerstone in many iterative workloads, such as sparse iterative solvers and large-scale graph analytics. In such workloads, performance is often dominated by repeated SpMV operations on sparse matrices with highly irregular sparsity pattern. For example, the core computation of PageRank \cite{Berkhin2005A,gleich2015pagerank} is iterative SpMV on a graph transition matrix. Therefore, accelerating SpMV---especially on real-world graphs exhibiting irregular sparsity pattern \cite{ashari2014fast}---can substantially benefit a broad class of iterative applications. 

GPU SpMV is typically limited by the memory system rather than arithmetic throughput \cite{cuthill1969reducing, lv2023survey}. For single precision, processing each nonzero incurs at least 8 bytes memory traffic (4 for matrix entry and 4 for input vector element), while the sustained memory bandwidth of GPUs remains far below their peak compute throughput. Consequently, the performance of GPU SpMV is primarily determined by (i) \emph{workload balance} and (ii) \emph{memory access efficiency}. Although optimal workload balance \cite{hu2024fastload} does not inherently guarantee high performance, imbalance can severely degrade the computational throughput. Once the workload balance is achieved, memory access efficiency becomes the dominant factor.

Existing GPU SpMV methods address these challenges from different directions. CSR5 \cite{liu2015csr5} partitions nonzeros into fixed-size tiles and encodes the execution flow using compact bit-field metadata, enabling reusable scheduling information after preprocessing. Merge-Based SpMV \cite{merrill2016merge} partitions the merge-path to obtain balanced workload segments and uses shared memory to coordinate coalesced matrix loading with parallel reduction. HOLA \cite{steinberger2017globally} further improves output updates by buffering partial sums in shared memory. However, these methods still leave room for improvement in repeated irregular SpMV: scheduling metadata generation, tile-boundary handling, and output committing may introduce auxiliary overheads, while special structures such as empty rows and long rows require dedicated treatment.

Iterative workloads provide an important opportunity for reducing such overheads. Since the sparsity pattern of the matrix remains unchanged across iterations, scheduling metadata can be generated once and reused many times. Consequently, lightweight preprocessing becomes practical when its cost can be amortized over repeated SpMV calls. This observation motivates a design that combines balanced merge-path partitioning, compact reusable metadata, coalesced memory access, and fused in-kernel handling of boundary and output updates.

Based on this design principle, we propose MERBIT, a GPU-based SpMV method designed for repeated irregular SpMV workloads. MERBIT combines the key ideas of CSR5 and Merge-Based SpMV: globally, it partitions the merge-path to obtain balanced work units; locally, it encodes the execution flow of each segment using compact bit-field descriptors. During SpMV, MERBIT uses shared memory to bridge coalesced matrix loading, parallel reduction, and coalesced output committing. It further introduces dedicated optimizations for long rows, metadata-free segmented reduction, and dual-buffer output committing.

The main contributions of this paper are summarized as follows.
\begin{enumerate}
	\item \textbf{MERBIT storage format}. We introduce a sparse matrix storage format which unifies the strengths of CSR5 and Merge-Based SpMV. MERBIT encodes the merge-path in compact bit-field representation, enabling workload balance through a one-time preprocessing.
	\item \textbf{MERBIT SpMV kernel.} We design a MERBIT SpMV kernel that integrates coalesced matrix loading, parallel reduction, and coalesced output committing through shared memory. The kernel further incorporates optimizations for degenerate tiles, long rows, segmented reduction, and dual-buffer output committing.
	\item \textbf{Extensive evaluation.}  We evaluate MERBIT on 50 large irregular datasets collected from the SuiteSparse Matrix Collection. The results show that MERBIT outperforms competitive baselines including the vendor library (NVIDIA cuSPARSE), open-source library (Ginkgo), and the academic approaches. Compared to cuSPARSE COO, MERBIT delivers an average of 1.27$\times$ and 1.25$\times$ speedups in single and double precision, respectively. 
\end{enumerate}

The rest of this paper is organized as follows. In Section 2, we review the related work. Section 3 introduces SpMV, focusing on CSR5 and Merge-Based SpMV. Section 4 presents MERBIT, including the storage format, kernel design, and the parameter optimization. Section 5 reports experimental results. We conclude this paper in Section 6.

\section{Related Work}
A wide range of sparse matrix storage formats and corresponding SpMV kernels have been proposed, including Coordinate (COO), Compressed Sparse Row (CSR), ELLPACK (ELL), and Diagonal (DIAG) (see \cite{bell2009implementing} for implementation and comparison of these kernels). Extensions such as Sliced-ELLPACK (SELL) \cite{monakov2010automatically}, CSR-Vector \cite{liu2015lightspmv} and Hybrid (HYB) are developed to further improve the performance for specific sparsity patterns. Additionally, cache blocking models \cite{vuduc2003automatic, temam1992characterizing}, kernel analysis techniques \cite{li2014performance, ashari2015model, grewe2011automatically, choi2010model} and parameter optimization methods \cite{maggioni2016optimization, baskaran2009optimizing, kaya2014analysis} are introduced to further enhance the performance. In practice, open-source libraries such as Ginkgo \cite{anzt2022ginkgo} and vendor libraries such as NVIDIA cuSPARSE \cite{cusparse_doc} provide highly optimized SpMV routines and are widely used as baselines in prior studies. 
\begin{table}[!t]
	\caption{Comparison of representative edge-partitioned GPU SpMV kernels.}
	\label{tab:kernel_compare}
	\centering
	\tiny
	\setlength{\tabcolsep}{3.5pt}
	\renewcommand{\arraystretch}{1.18}
	\begin{tabularx}{\textwidth}{@{}p{2.25cm}YYYY@{}}
		\toprule
		 &
		\textbf{\makecell{Merge-Based\\SpMV}} &
		\textbf{CSR5} &
		\textbf{HOLA} &
		\textbf{MERBIT} \\
		\midrule
		
		Workload partitioning
		& merge-path segments
		& nonzero-based tiles
		& nonzero-based tiles
		& merge-path segments \\
		
		Matrix loading
		& shared-memory staged, coalesced
		& tile-wise transposed, coalesced
		& warp-level transposed, partially coalesced
		& shared-memory staged, coalesced \\
		
		Output update
		& arbitrary global writes
		& arbitrary global writes
		& buffered and coalesced
		& buffered and coalesced \\
		
		Empty-row handling
		& naturally handled by merge path
		& special-case handling
		& special-case handling
		& naturally handled by merge path \\
		
		Long-row handling
		& no dedicated path
		& warp-level reduction
		& warp-level reduction
		& warp-level fast path \\
		
		Boundary handling
		& separate fix-up
		& separate fix-up
		& atomic
		& atomic \\
		
		\bottomrule
	\end{tabularx}
\end{table}

From a graph perspective, workload balance relates to balanced graph partitioning. Vertex-partitioning approaches such as CSR and ELL assign rows to processing units (thread, warp or block) and work well for matrices exhibiting regular sparsity patterns. However, the workload of SpMV is proportional to the number of nonzeros; matrices derived from web graphs and social networks often exhibit highly irregular sparsity, leading to severe imbalance. This motivates edge-partitioned approaches, including CSR5, HOLA, and Merge-Based SpMV. Table~\ref{tab:kernel_compare} compares these representative GPU SpMV kernels along the design dimensions that are most relevant to repeated irregular SpMV.

CSR5 partitions nonzeros into fixed-size two-dimensional tiles and assigns one warp to each tile. It uses compact bit-field descriptors to encode the execution flow, and it applies tile-wise transpose to improve matrix-access coalescing. However, CSR5 still requires tile-boundary handling, and its partial sums may be written to arbitrary output locations. The load--compute progression within a tile can also introduce long dependency chains.

HOLA adopts a similar edge-partitioning strategy, but constructs the scheduling metadata (block and thread starting coordinates) at run time using an additional kernel. It bridges coalesced matrix accesses with parallel reduction by a warp-level transpose implementation. Nevertheless, its multiplications and warp-level transpose require frequent synchronizations, incurring substantial warp barrier stall, especially in higher precision. Consequently, HOLA delivers a relatively unstable cross-precision performance. 

Merge-Based SpMV provides a hybrid partitioning strategy which divides the merge-path into fixed-size tiles. The scheduling metadata (starting coordinates of lanes along the merge-path) is obtained by merge-search. Modern implementations can coalesce the matrix accesses via shared memory. However, it lacks optimizations for tiles containing only one long row; the partial sums are stored in a non-coalesced manner; and the separate kernels for generating scheduling metadata and tile-boundary fix-up can introduce auxiliary overheads.

SpMV is the cornerstone in many iterative workloads, including PageRank, Arnoldi iteration, and Krylov subspace methods. In such settings, the sparsity structure of the matrix remains unchanged, enabling reusable scheduling metadata constructed by one-time preprocessing. Therefore, an important research direction is to retain the benefits of workload balance and coalesced memory accesses, while reducing the per-iteration auxiliary overhead by fusing multiple steps into the main multiplication kernel.

Beyond SpMV, there also exist GPU graph analytic frameworks that provide end-to-end implementations of PageRank and other graph workloads (e.g., Gunrock \cite{wang2016gunrock}). These frameworks optimize the overall pipeline and include non-SpMV components such as graph representation, preprocessing and convergence checks. In contrast, this work focuses on SpMV and its per-iteration overhead. Therefore, a full framework-level comparison is outside the scope of this paper.

\section{Preliminary}
In this section, we review the SpMV formulation, and two key building blocks related to our design: (i) compact bit-field representation in CSR5 and (ii) merge-path in Merge-Based SpMV. 

\subsection{SpMV}
Let $\bm{A}\in\mathbb{R}^{n_{1}\times n_{2}}$ be a sparse matrix with \textit{m} nonzero entries, $\bm{x}\in\mathbb{R}^{n_{2}}$ an input vector, and $\bm{y}\in\mathbb{R}^{n_{1}}$ an output vector. SpMV is defined as
\[
	\bm{y}=\bm{A}\bm{x}.  
\]
In this paper, matrices typically relate to graphs; hence, $\bm{A}$ is usually square ($n_{1}=n_{2}=n$). We adopt the square-matrix notation when it simplifies the exposition, although the proposed method does not require $\bm{A}$ to be square.

A widely used storage format is CSR, which represents $\bm{A}$ using three arrays: \textit{offset} of length \textit{n}, \textit{col\_ind} and \textit{values} of length \textit{m}. For matrices exhibiting irregular sparsity pattern, CSR-Based SpMV can incur severe workload imbalance and non-coalesced matrix access. Thus, a wide range of 2D-tiling methods are proposed.
\subsection{CSR5}
\begin{figure}[h]
	\centering
	\includegraphics[width=0.90 \textwidth]{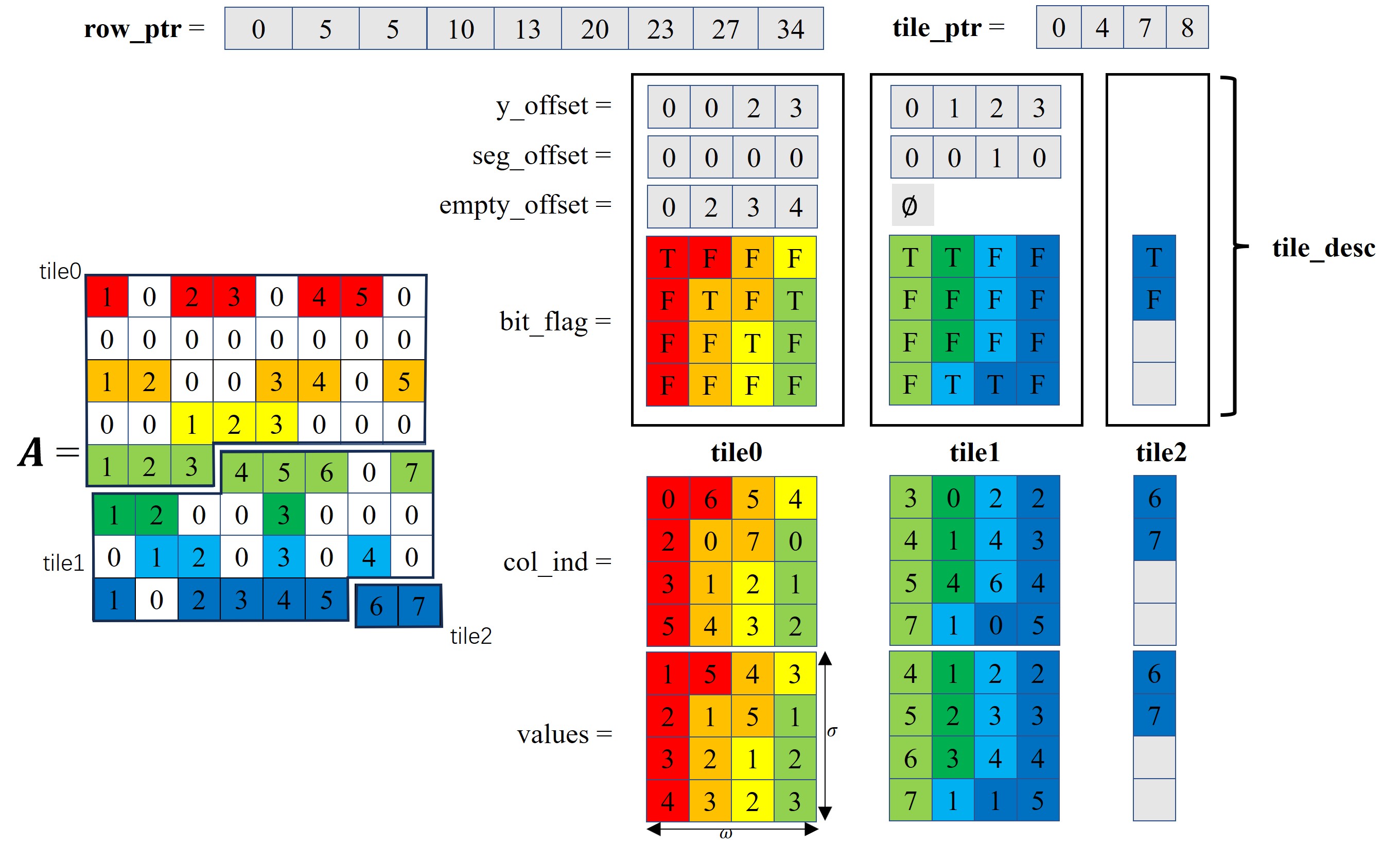}
	\caption{CSR5 storage format ($\omega=4$, $\sigma=4$ )}
	\label{csr5}
\end{figure}

CSR5 partitions nonzeros into 2D tiles of size $\omega \times \sigma$. Figure~\ref{csr5} illustrates an example with $\omega=4$ and $\sigma=4$, the matrix $\bm{A}$ is partitioned into three tiles, where tile0 and tile1 are full tiles with 16 nonzeros each, while tile2 is an incomplete tile containing the remaining 2 nonzeros. 

Each tile $\textit{i}$ maintains a pointer $\textit{tile\_ptr}[i]$ recording the row index of the first nonzero in this tile. In Figure~\ref{csr5}, $\textit{tile\_ptr} = [0, 4, 7, 8]$ means tile0, tile1 and tile2 start from rows 0, 4 and 7, respectively; the last entry 8 indicates the end row index.
\begin{figure}[h]
	\centering
	\footnotesize
	$\underbrace{31-15}_{\textit{bit\_flag}} \underbrace{14-10}_{\textit{seg\_offset}} \underbrace{9-0}_{\textit{y\_offset}}$
	\caption{The structure of CSR5 lane descriptor ($\omega = 32$, $\sigma = 17$)}
	\label{csr5lanedesc}
\end{figure}

For each tile $\textit{i}$, CSR5 further maintains lane descriptors $\textit{tile\_desc}[i][j]$ for \textit{j} $\in [0, \omega-1]$, consisting of 
\begin{enumerate}
	\item [$\bullet$] row offset ($\textit{y\_offset}[i][j]$) for start row of this lane;
	\item [$\bullet$] segment offset ($\textit{seg\_offset}[i][j]$) for segmented reduction;
	\item [$\bullet$] row boundary flags ($\textit{bit\_flag}[i][j]$).
\end{enumerate}
Hence, the starting row index of the $j$-th lane is 
\[
	\textit{tile\_ptr}[i] + \textit{y\_offset}[i][j].
\]
In addition, $\textit{empty\_offset}$ is utilized to find the correct locations of partial sums in output vector when tile contains empty rows. In tile0, $\textit{y\_offset}=[0,0,2,3]$, $\textit{seg\_offset}=[0,0,0,0]$ and $\textit{empty\_offset}=[0,2,3,4]$; for lane 1, \textit{bit}\_\textit{flag} = [\texttt{F}, \texttt{T}, \texttt{F}, \texttt{F}] means the second entry of this lane starts a new row at row
\[
	\textit{empty\_offset}[\textit{tile\_ptr}[0] + \textit{y\_offset}[0][1]] = 2.
\] 

As depicted in Figure~\ref{csr5lanedesc}, $\textit{tile\_desc}$ is encoded into compact bit-field representation. For each lane $j$ in tile $i$, $\textit{y\_offset}[i][j]$, $\textit{seg\_offset}[i][j]$ and $\textit{bit\_flag}[i][j]$ are integrated into a 32-bit integer, providing reusable scheduling metadata with small memory footprint. The CSR5-Based SpMV kernel is outlined in Appendix~A, Algorithm~1. 

\subsection{Merge-Based SpMV}
\begin{figure}[h]
	\centering
	\includegraphics[width=0.96 \textwidth]{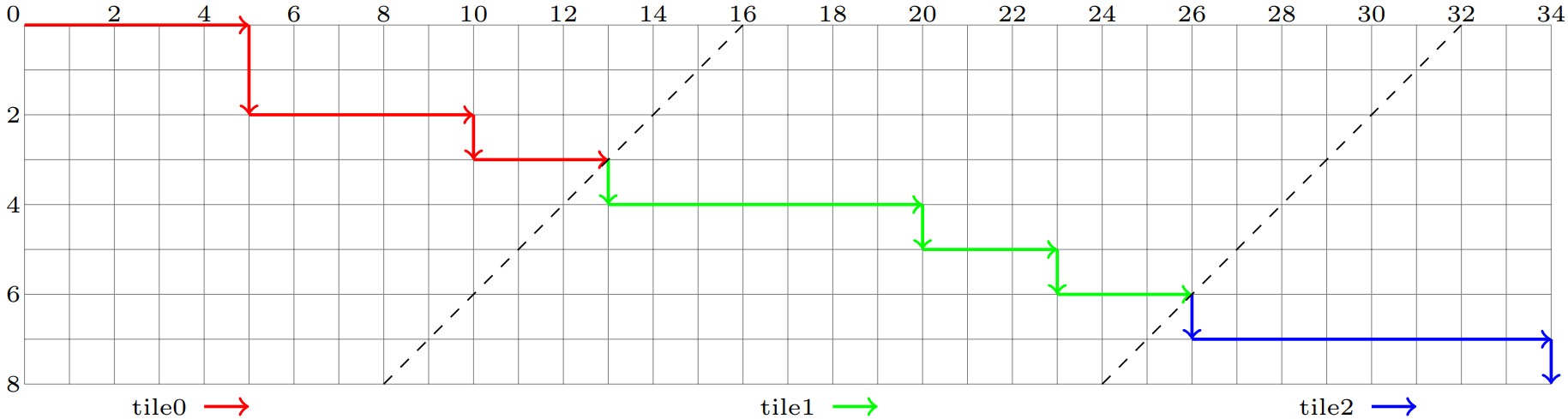}
	\caption{The merge-path of $\bm{A}$ ($\omega = 4$, $\sigma = 4$)}
	\label{mergepath}
\end{figure}

Merge-Based SpMV is built upon merge-path---a monotone path from $(0,0)$ to $(m,n)$ on the 2D grid. As depicted in Figure \ref{mergepath}, the merge-path of $\bm{A}$ in Figure \ref{csr5} is the colored path from $(0,0)$ to $(34,8)$, where the \textit{x}-coordinate denotes the index of the nonzero stream (\textit{col}\_\textit{ind} and \textit{values}), while \textit{y}-coordinate denotes the row index. A step starting from $(i,j)$ and moving rightward indicates accumulating $\textit{values}[i] \times \textit{x}[\textit{col\_ind}[i]]$ into the partial sum of $y_{j}$, while a downward step represents storing the partial sum into $y_{j}$. 
\begin{algorithm}[h]
	\scriptsize
	\caption{ \textbf{merge\_search}}
	\begin{algorithmic}[1]
		\Require{ \textit{offset}, \ \textit{diag}, \ \textit{n}, \ \textit{m}. }
		\Ensure{\textit{coord\underline{~}x}, \ \textit{coord\underline{~}y}. }
		\State{$\textit{y\_min} = \max(\textit{diag} - \textit{m}, \ 0)$;}
		\State{$\textit{y\_max} = \min(\textit{diag}, \ \textit{n})$;}
		\While{$\textit{y\_min} < \textit{y\_max}$}
		\State{$\textit{mid} = (\textit{y\_min} + \textit{y\_max}) >> 1$;}
		\If{$\textit{offset}[\textit{mid} + 1] \leq \textit{diag} - \textit{mid} - 1$}
		\State{$\textit{y\_min} = \textit{mid} + 1$;}
		\Else
		\State{$\textit{y\_max} = \textit{mid}$;}
		\EndIf
		\EndWhile
		\State{$\textit{coord\_x} = \textit{diag} - \textit{y\_min}$;}
		\State{$\textit{coord\_y} = \min(\textit{y\_min}, \ \textit{n})$;}
	\end{algorithmic}
	\label{mergesearch}
\end{algorithm}

Merge-path is obtained by \texttt{merge\_search}. As demonstrated in Algorithm~\ref{mergesearch}, given the location of merge-path (\textit{diag}), it performs binary search along this diagonal
\[
	i + j = \textit{diag}, 
\]
starting from $(\textit{diag}, 0)$, and returns the coordinate $(i,j)$ of the intersection point with merge-path. 

Merge-Based SpMV partitions the merge-path into 2D tiles of size $\omega\times\sigma$, assigns one warp to each tile and maps $\omega$ threads to the $\omega$ lanes. As depicted in Figure~\ref{mergepath}, the merge-path is partitioned into three tiles, where tile0 and tile1 are full tiles with 16 steps each, while tile2 is an incomplete one containing the remaining 10 steps. In tile0, lane 0-3 consume the segments $(0,0)$-$(4,0)$, $(4,0)$-$(6,2)$, $(6,2)$-$(10,2)$ and $(10,2)$-$(13,3)$, respectively. By treating right step and down step as unit work, Merge-Based SpMV achieves a balanced workload partitioning. Moreover, empty rows naturally appear as consecutive down steps on the merge-path and can be handled without auxiliary empty-row correction array. The Merge-Based SpMV kernel is outlined in Appendix~B, Algorithm~3.

\section{MERBIT}
This section presents MERBIT, a GPU-based SpMV method designed for repeated irregular workloads. MERBIT first constructs reusable merge-path metadata during preprocessing, and then uses the metadata to guide an efficient GPU SpMV kernel. The design goal is to retain the workload balance of merge-path partitioning while reducing the per-iteration overhead of repeated scheduling, boundary handling, and output committing.

\subsection{MERBIT Storage Format}
\subsubsection{Data Layout}
\begin{figure}[htpb]
	\centering
	\includegraphics[width=0.90 \textwidth]{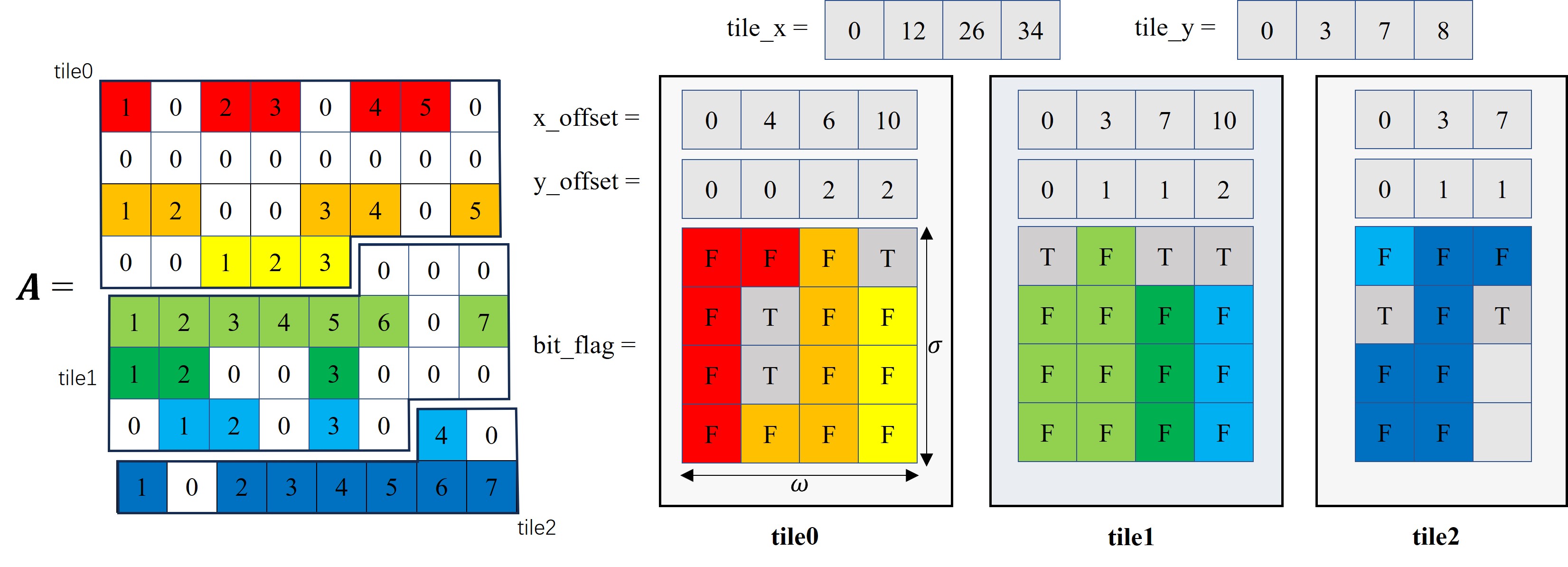}
	\caption{MERBIT storage format ($\omega=4$, $\sigma=4$ )}
	\label{mergebit}
\end{figure}
\begin{figure}[htpb]
	\centering
	\footnotesize
	$\underbrace{31-18}_{\textit{bit\_flag}} \underbrace{17-9}_{\textit{y\_offset}} \underbrace{8-0}_{\textit{x\_offset}}$
	\caption{The structure of MERBIT lane descriptor ($\omega = 32$ and $\sigma = 14$)}
	\label{mergebitlanedesc}
\end{figure}
MERBIT partitions the merge-path into tiles of size $\omega\times\sigma$, where $\omega$ is the number of lanes per tile (typically equal to the warp size on NVIDIA GPU) and $\sigma$ is the number of steps contained in each lane. Thus, MERBIT generates 
\[
	\textit{tile\_num} = \Big \lceil \frac{m+n}{\omega \sigma} \Big \rceil
\]
tiles and 
\[
	\textit{lane\_num} = \Big \lceil \frac{m+n}{\sigma} \Big \rceil
\]
lanes in total. MERBIT represents the merge-path using five arrays: $\textit{tile\_x}$, $\textit{tile\_y}$, $\textit{x\_offset}$, $\textit{y\_offset}$ and $\textit{bit\_flag}$.
\begin{itemize}
	\item $\textit{tile\_x}$ and $\textit{tile\_y}$ are arrays of length $\textit{tile\_num} + 1$, $(\textit{tile\_x}[i], \textit{tile\_y}[i])$ records the starting coordinate of tile \textit{i} on the merge-path, with the last entry being $(m,n)$.  
	\item $\textit{x\_offset}$ and $\textit{y\_offset}$ are arrays of length $\textit{lane\_num}$. For the $j$-th lane of the $i$-th tile, $(\textit{x\_offset}[i][j], \textit{y\_offset}[i][j])$ records the lane's starting-coordinate offset relative to the starting coordinate of its tile.
	\item $\textit{bit\_flag}$ is of length $\textit{lane\_num}$, $\textit{bit\_flag}[i][j]$ encodes the $\sigma$ steps of $j$-th lane into bit-flags, where a bit value \texttt{T} marks a down step and a bit value \texttt{F} marks a right step.
\end{itemize}
Therefore, the coordinate of the $k$-th step in lane $j$ of tile $i$ is
\[
	(\textit{tile\_x}[i] + \textit{x\_offset}[i][j] + k_x, \ \textit{tile\_y}[i] + \textit{y\_offset}[i][j] + k_y),
\]
where $k_x$ and $k_y$ are the numbers of right steps and down steps within the first $k$ bits of $\textit{bit\_flag}[i][j]$, respectively.

Figure~\ref{mergebit} illustrates an example for $\bm{A}$ in Figure~\ref{csr5} with $\omega = 4$ and $\sigma = 4$. Three tiles are generated with starting coordinates $(0,0)$, $(12,3)$ and $(26,7)$, hence, $\textit{tile\_x} = [0,12,26,34]$ and $\textit{tile\_y} = [0,3,7,8]$. For tile1, the starting coordinates of its four lanes are $(12, 3)$, $(15, 4)$, $(19,4)$ and $(22, 5)$, hence, the corresponding offsets are $\textit{x\_offset} = [0, 3, 7, 10]$ and $\textit{y\_offset} = [0, 1, 1, 2]$. The first lane in tile1 initially finalizes row 3 (down step) and then performs three right steps for row 4, hence, the corresponding $\textit{bit\_flag} = \texttt{TFFF}$. 

The five arrays constitute the storage format. To reduce the memory footprint, we further encode $\textit{x\_offset}[i][j]$, $\textit{y\_offset}[i][j]$ and $\textit{bit\_flag}[i][j]$ into compact bit-field representation:  
\begin{itemize}
	\item $\sigma$ bits are sufficient for $\textit{bit\_flag}[j]$, as the height of each lane does not exceed $\sigma$;  
	\item $\lceil\log_{2}{(\omega \sigma)}\rceil$ bits are sufficient for $\textit{x\_offset}[j]$, as each tile contains at most $\omega \times \sigma$ nonzeros;
	\item $\lceil\log_{2}{(\omega \sigma)}\rceil$ bits are sufficient for $\textit{y\_offset}[j]$, as each tile involves at most $\omega \times \sigma$ rows.   
\end{itemize}
Consequently, it is feasible to pack $\textit{x\_offset}[i][j]$, $\textit{y\_offset}[i][j]$ and $\textit{bit\_flag}[i][j]$ into a 32-bit lane descriptor when 
\[
	2\lceil\log_{2}{(\omega \sigma)}\rceil + \sigma \leq 32.
\]
As illustrated in Figure~\ref{mergebitlanedesc}, for $\omega=32$ and $\sigma=14$, the above condition holds, and we integrate $\textit{x\_offset}$, $\textit{y\_offset}$ and $\textit{bit\_flag}$ into 32-bit lane descriptors ($\textit{lane\_desc}$). As a result, MERBIT stores the scheduling metadata
\[
	\textit{TILE} = \{\textit{tile\_x}, \ \textit{tile\_y}, \ \textit{lane\_desc}\}.
\]
\subsubsection{Generation of \textit{TILE}}
As demonstrated in Algorithm \ref{mergebitprepocess}, \textit{TILE} is generated by a GPU kernel, with each warp addressing one tile and each thread handling one lane. The whole procedure consists of three steps.  

\textit{(1) Locating tile starts and lane offsets.} For the $j$-th lane, the corresponding diagonal position on the merge-path is
computed as
\[
\textit{diag} = j \cdot \sigma .
\]
Then, \texttt{merge\_search} maps this one-dimensional position to the merge-path coordinate $(\textit{coord\_x}, \textit{coord\_y})$. For a warp, lane 0 corresponds to the first lane of the current tile, so its coordinate is also the starting coordinate of the tile. This coordinate is broadcast to the other lanes in the same warp through warp-level shuffle operations. Each thread then computes the relative offset of its lane start with respect to the tile start, namely \textit{x\_offset} and \textit{y\_offset}. In this way, the tile-level global coordinate and the lane-level local offsets are generated simultaneously.
\begin{algorithm}[h]
	\scriptsize
	\caption{ \textbf{generate\_tile} }
	\begin{algorithmic}[1]
		\Require{$\textit{offset}, \ \omega, \ \sigma, \ \textit{n}, \ \textit{m}, \ \textit{tile\_num}, \ \textit{lane\_num}$. }
		\Ensure{$\textit{tile\_x}$, \ \textit{tile\_y}, \ \textit{lane\_desc}.}
		\State{$\textit{j} \leftarrow \textit{blockDim.x} \cdot \textit{blockIdx.x} + \textit{threadIdx.x}$;}
		\State{$\textit{lane\_nnz} \leftarrow ((\textit{j} + 1) \cdot \sigma < \textit{m} + \textit{n})?\quad \sigma : (\textit{m} + \textit{n} - \textit{j} \cdot \sigma)$;}
		\State{\textit{lid} $\leftarrow \textit{j} \bmod \omega$;}
		\State{$\textit{i} \leftarrow \lfloor \frac{j}{\omega} \rfloor$;}
		\If{\textit{j} $<$ \textit{lane\_num}}
		\State{$\textit{diag} \leftarrow \textit{j} \cdot \sigma$;}
		\State{$(\textit{coord\_x}, \textit{coord\_y}) \leftarrow \text{merge\_search}(\textit{offset}, \ \textit{diag})$;} 
		\State{$\textit{warp\_x\_start} \leftarrow \texttt{shfl\_sync}(\textit{coord\_x}, 0)$;}
		\State{$\textit{warp\_y\_start} \leftarrow \texttt{shfl\_sync}(\textit{coord\_y}, 0)$;}
		\State{$\textit{x\_offset} \leftarrow \textit{coord\_x} - \textit{warp\_x\_start}$;}
		\State{$\textit{y\_offset} \leftarrow \textit{coord\_y} - \textit{warp\_y\_start}$;}
		\State{$\textit{bit\_flag} \leftarrow \text{0x0}$;}
		\For{$\textit{idx}=0$ to $\textit{lane\_nnz} - 1$;}
		\If{$\textit{coord\_x} < \textit{offset}[\textit{coord\_y}+1]$}
		\State{$++\textit{coord\_x}$;}
		\Else
		\State{\textit{bit\_flag} $\vert=$ (0x1 $<<$ \textit{idx});}
		\State{$++\textit{coord\_y}$;}
		\EndIf
		\EndFor
		\State{\textit{lane\_desc}[\textit{j}] $\leftarrow$ (\textit{bit\_flag} $<<$ 18) $\vert$ (\textit{y\_offset} $<<$ 9) $\vert$ \textit{x\_offset};} 
		\If{\texttt{all\_sync}(\textit{coord\_y} == \textit{warp\_y\_start})}
		\State{\textit{warp\_y\_start} $\leftarrow$ 0x80000000 $\vert$ \textit{warp\_y\_start};}
		\EndIf
		\If{\textit{lid} == 0}
		\State{\textit{tile\_x}[\textit{i}] $\leftarrow$ \textit{warp\_x\_start};} 
		\State{\textit{tile\_y}[\textit{i}] $\leftarrow$ \textit{warp\_y\_start};} 
		\EndIf
		\ElsIf{\textit{j} == \textit{lane\_num}}
		\State{\textit{tile\_x}[\textit{tile\_num}] $\leftarrow$ \textit{m};} 
		\State{\textit{tile\_y}[\textit{tile\_num}] $\leftarrow$ \textit{n};} 
		\EndIf
	\end{algorithmic}
	\label{mergebitprepocess}
\end{algorithm}

\textit{(2) Generating the lane direction flags.} After obtaining the starting coordinate of a lane, each thread simulates at most $\sigma$ steps along the merge-path to generate the direction flag \textit{bit\_flag}. At each step, the current coordinate $(\textit{coord\_x}, \textit{coord\_y})$ is compared with the next row boundary $\textit{offset}[\textit{coord\_y}+1]$. If
\[
	\textit{coord\_x} < \textit{offset}[\textit{coord\_y}+1],
\]
the current step is a right step, meaning that the next nonzero still belongs to the current row; the thread therefore increments \textit{coord\_x}. Otherwise, the current row is completed and the path moves downward to the next row. In this case, the corresponding bit in \textit{bit\_flag} is set, and \textit{coord\_y} is incremented. For the tail lane, whose remaining path length may be smaller than $\sigma$, only the valid number of steps is simulated.

\textit{(3) Packing descriptors and marking special tiles.} Once the lane-local offsets and direction flags have been generated, each thread packs \textit{x\_offset}, \textit{y\_offset}, and \textit{bit\_flag} into a 32-bit \textit{lane\_desc}. Thread 0 of each warp writes the tile starting coordinate to \textit{tile\_x} and \textit{tile\_y}. The last entry of the tile-coordinate arrays is set to $(m,n)$, which marks the end of the merge-path. MERBIT also identifies long-row tiles during preprocessing. If all lanes in a tile remain in the same row after simulating their merge-path segments, the tile contains only one row. Such a tile is marked by setting the most significant bit of its \textit{tile\_y} entry. This mark allows the SpMV kernel to bypass descriptor parsing and use a long-row fast path.

Overall, the preprocessing cost is
\[
	O(\textit{lane\_num} \cdot (\log n + \sigma)),
\]
where the $\log n$ term comes from \texttt{merge\_search} and the $\sigma$ term comes from lane-local path simulation. Since TILE only records reusable scheduling metadata, its one-time construction cost can be amortized over many SpMV iterations.

\subsection{MERBIT-Based SpMV Kernel}
This subsection describes how MERBIT uses the encoded merge-path schedule to execute SpMV efficiently on GPU. The kernel is designed around three principles. First, the merge-path schedule assigns each lane a fixed-length
segment, so the workload is balanced over both nonzero products and row boundary events. Second, the contiguous matrix stream within each tile is staged in shared memory, which improves the coalescing of matrix-value and
column-index accesses before local accumulation. Third, partial sums are buffered at the block level and committed cooperatively, so that most output writes are contiguous and only tile or block boundary rows require atomic accumulation.
\begin{algorithm}[h]
	\scriptsize
	\caption{\textbf{MERBIT SpMV}}\label{mergebitspmv}
	\begin{algorithmic}[1]
		\Require{\textit{tile\_x}, \ \textit{tile\_y}, \ \textit{lane\_desc}, \ \textit{m}, \ \textit{n}, \ $\omega$, \ $\sigma$, \ \textit{col\_ind}, \ \textit{values}, \ \textit{x}}
		\Ensure{\textit{y},\ \textit{z}}
		\State{$\textit{j} \leftarrow \textit{blockDim.x} \cdot \textit{blockIdx.x} + \textit{threadIdx.x}$;}
		\State{$\textit{tid} \leftarrow \textit{threadIdx.x}$;}
		\State{$\textit{lid} \leftarrow \textit{j} \bmod \omega$;}
		\State{$\textit{i} \leftarrow \lfloor \textit{j}/\omega \rfloor$;}
		\State{$\textit{lane\_nnz} \leftarrow ((\textit{j} + 1) \cdot \sigma < \textit{m} + \textit{n})? \ \sigma : (\textit{m} + \textit{n} - \textit{j} \cdot \sigma)$;}
		\State{$\textit{bs} \leftarrow \lfloor (\textit{blockIdx.x}\cdot\textit{blockDim.x})/\omega \rfloor$;}
		\State{$\textit{be} \leftarrow \lfloor ((\textit{blockIdx.x} + 1) \cdot\textit{blockDim.x})/\omega \rfloor$;}
		\State{$y_{bs} \leftarrow \text{ROW}(\textit{tile\_y}[\textit{bs}])$;} \Comment{$\text{ROW}(\cdot)$ returns the row index}
		\State{$y_{be} \leftarrow \text{ROW}(\textit{tile\_y}[\textit{be}])$;} \Comment{$\text{ROW}(\cdot)$ returns the row index}
		\State{$y_{se} \leftarrow y_{be} - y_{bs}$;}
		\State{$x_{bs} \leftarrow \textit{tile\_x}[\textit{bs}]$;} 
		\State{$x_{ws} \leftarrow \textit{tile\_x}[\textit{i}]$;}
		\State{$x_{we} \leftarrow \textit{tile\_x}[\textit{i}+1]$;}
		\State{$x_{se} \leftarrow x_{we} - x_{ws}$;}
		\State{$x_{bw} \leftarrow x_{ws} - x_{bs}$;}
		\If{$x_{ws} < x_{we}$} 
		\State{$y_{ws} \leftarrow \text{ROW}(\textit{tile\_y}[\textit{i}])$;}
		\State{$y_{bw} \leftarrow y_{ws} - y_{bs}$;}
		\If{\text{MSB}(\textit{tile\_y}[\textit{i}])}
		\State{$\texttt{tile\_fast\_tackle}(\textit{values}[x_{ws}, x_{we}), \ \textit{col\_ind}[x_{ws}, x_{we}), \ \textit{x}, \ \&\textit{s\_y}[y_{bw}]$)};
		\Else
		\State{$\texttt{load\_smem}(\textit{values}[x_{ws}, x_{we}), \ \textit{col\_ind}[x_{ws}, x_{we}), \ x_{se}, \ \textit{lid}, \ \textit{x}, \ \&\textit{s\_data}[x_{bw}]$)};
		\State{\texttt{syncwarp}()};
		\State{$\texttt{tile\_normal\_tackle}(\textit{s\_data}[x_{bw}], \ \textit{lane\_desc}[\textit{j}], \ \textit{lane\_nnz}, \ \textit{lid}, \ \&\textit{s\_y}[y_{bw}]$)};
		\EndIf
		\EndIf
		\State{\texttt{syncthreads}()};
		\State{\texttt{load\_mem}($\textit{s\_y}, \ y_{se}, \ \&y[y_{bs}], \ \&z[y_{bs}]$)};
	\end{algorithmic}
\end{algorithm}

Algorithm \ref{mergebitspmv} shows the main control flow of MERBIT-Based SpMV kernel. Each warp is responsible for one tile and each thread handles one lane. For a tile $i$, the nonzero range is determined by
$x_{ws} = \textit{tile\_x}[i]$ and $x_{we} = \textit{tile\_x}[i+1]$. If $x_{ws} = x_{we}$, the tile contains no nonzero and can be skipped directly. Otherwise, a regular tile is processed in three stages.

First, the warp computes the products $\textit{values}\cdot x$ over the range $[x_{ws},x_{we})$ and stores them into shared memory. This staging step keeps the accesses to \textit{values} and \textit{col\_ind} contiguous across the warp, while the indirect accesses to the input vector are handled during product generation (line~22; see Appendix~C, Algorithm~5).

Second, after a warp-level synchronization, each thread parses its lane descriptor to recover the lane-local offsets and the row-boundary bit flags. The thread then scans the encoded merge-path steps. A right step consumes
one staged product and adds it to the current partial sum, whereas a down step finalizes the current row segment and writes the partial sum into the shared-memory output buffer (line~24; see Appendix~C, Algorithm \ref{tilenormaltackle}).

Third, after all warps in a thread block have finished processing their tiles, a block-level synchronization ensures that all partial sums have been written to the shared-memory output buffer. The block then cooperatively commits the buffered results to global memory. (line~28; see Appendix~C, Algorithm~6).
\begin{algorithm}[h]
	\scriptsize
	\caption{ \textbf{tile\_normal\_tackle}}
	\begin{algorithmic}[1]
		\Require{ \textit{s\_data}, \ \textit{lane\_desc}, \ \textit{lane\_nnz}, \ \textit{lid}, \ \textit{s\_y}. }
		\Ensure{ \textit{s\_y}.}
		\State{$\textit{x\_offset} \leftarrow \text{X\_OFFSET}(\textit{lane\_desc})$;} \Comment{$\text{X\_OFFSET}(\cdot)$ returns \textit{x\_offset}}
		\State{$\textit{y\_offset} \leftarrow \text{Y\_OFFSET}(\textit{lane\_desc})$;} \Comment{$\text{Y\_OFFSET}(\cdot)$ returns \textit{y\_offset}}
		\State{$\textit{bit\_flag} \leftarrow \text{BIT\_FLAG}(\textit{lane\_desc})$;} \Comment{$\text{BIT\_FLAG}(\cdot)$ returns \textit{bit\_flag}}
		\State{$\textit{first\_flag} \leftarrow \textit{true}$;} 
		\State{$\textit{sum} \leftarrow 0$;} 
		\State{$\textit{lane\_flag} \leftarrow (\textit{lid} == 0)? \ \textit{true} : \textit{false}$;}
		\For {$k=0$ to $\textit{lane\_nnz} - 1$;}
		\State{\textit{flag} $\leftarrow$ 0x1 \& (\textit{bit\_flag} $>>$ \textit{k})}
		\State{$\textit{sum} \ += (1 - \textit{flag}) \cdot \textit{s\_data}[\textit{x\_offset}]$;}
		\If{\textit{flag}}
		\If{$\textit{first\_flag}$}
		\State{atomicAdd(\&\textit{s\_y}[\textit{y\_offset}], \textit{sum});}
		\State{\textit{first\_flag} $\leftarrow$ \textit{false};}
		\Else
		\State{\textit{s\_y}[\textit{y\_offset}] $\leftarrow$ \textit{sum};}
		\EndIf
		\State{\textit{sum} $\leftarrow$ 0;}
		\EndIf
		\State{$\textit{x\_offset} \ += 1 - \textit{flag}$;}
		\State{$\textit{y\_offset} \ += \textit{flag}$;}
		\State{$\textit{lane\_flag} \ \vert= \textit{flag}$;}
		\EndFor
		\State{\texttt{WarpSegmentedSum}(\textit{sum}, \ \textit{y\_offset}, \ \textit{lane\_flag}, \ \textit{lid}, \&\textit{s\_y});}
	\end{algorithmic}
	\label{tilenormaltackle}
\end{algorithm}

Since each block contains at most $b$ lanes and each lane spans at most $\sigma$ steps, the segment length is upper bounded by $b\sigma$ steps, which corresponds to at most $b\sigma+1$ coordinates along the merge-path. Let $m_b$ and $n_b$ denote the number of right steps and down steps within this block segment, respectively, then we have $m_b + n_b \le b\sigma + 1$. In our implementation, we store the $m_b$ products in the first $m_b$ contiguous locations, and the $n_b$ partial sums in the subsequent $n_b$ locations. This yields a uniform and predictable shared memory requirement across blocks for a given $(b,\sigma)$, facilitating stable occupancy control. While shared memory accesses may incur bank conflicts due to irregular matrix sparsity pattern, this cost is typically preferable to the uncoalesced global memory accesses.
\begin{algorithm}[h]
	\scriptsize
	\caption{ \textbf{WarpSegmentedSum} } \label{warpsegmentedsum}
	\begin{algorithmic}[1]
		\Require{ \textit{sum}, \ \textit{y\_offset}, \ \textit{lane\_flag}, \ \textit{lid}, \ \textit{s\_y}. }
		\Ensure{\textit{s\_y}.}
		\State{$\textit{local\_sum} \leftarrow \textit{sum}$;}
		\State{$\textit{tmp\_sum} \leftarrow 0$;}
		\State{$\textit{offset} \leftarrow \text{0x1}$;}
		\State{$\textit{local\_flag} \leftarrow \textit{lane\_flag}$;}
		\State{$\textit{tmp\_flag} \leftarrow \textit{lane\_flag}$;}
		\While{\texttt{any\_sync}($\neg$\textit{local\_flag})}
		\State{$\textit{tmp\_sum} \leftarrow \texttt{shfl\_up\_sync}(\textit{local\_sum}, \ \textit{offset})$;}
		\State{\textit{tmp\_flag} $\leftarrow$ \texttt{shfl\_up\_sync}(\textit{local\_flag}, \ \textit{offset});}
		\State{\textit{local\_sum} += ($\neg$\textit{local\_flag} \& \textit{lid} $\geq$ \textit{offset})? \textit{tmp\_sum} : 0;}
		\State{\textit{local\_flag} $\leftarrow$ (\textit{local\_flag})? \textit{local\_flag} : \textit{tmp\_flag};}
		\State{\textit{offset} $<<$= 1;}
		\EndWhile
		\State{$\textit{src\_lane} \leftarrow (\textit{lid} == 0)? \ \omega - 1 : \textit{lid} - 1$;}
		\State{$\textit{row\_idx} \leftarrow \texttt{shfl\_sync}(\textit{y\_offset}, \ \textit{src\_lane})$;}
		\State{$\textit{tmp\_sum} \leftarrow \texttt{shfl\_sync}(\textit{local\_sum}, \ \textit{src\_lane})$;}
		\If{\textit{lane\_flag}}
		\State{atomicAdd(\&\textit{s\_y}[\textit{row\_idx}], \textit{tmp\_sum});}
		\EndIf
	\end{algorithmic}
\end{algorithm}

Within each lane, row boundaries are directly identified from $\textit{bit\_flag}$. However, the last partial sum of one lane may belong to the same row as the first partial sum of a neighboring lane. If every lane boundary were handled by atomic operations, the number of atomic additions would increase and the benefit of local buffering would be weakened. MERBIT therefore applies a metadata-free warp-level segmented reduction, shown in Algorithm~\ref{warpsegmentedsum}. The reduction propagates partial sums across lanes together with dynamically generated row-termination flags. In this way, MERBIT combines adjacent lane-local partial sums belonging to the same row without storing an additional \textit{segment\_offset} array. Compared with the CSR5-style segmented reduction, this design keeps the lane descriptor more compact while still reducing atomic additions at lane boundaries.

Moreover, MERBIT incorporates three optimizations to enhance the performance:

\textit{(1) Fast tackling for long rows.} For tiles that touch only one row, shared memory loading and row boundary identification is redundant. MERBIT marks such tiles during preprocessing (MSB in $\textit{tile\_y}$), and uses warp-level reduction to compute the tile partial sum (Algorithm~\ref{mergebitspmv} line~20), avoiding descriptor parsing and minimizing shared memory traffic.

\textit{(2) Warp-level segmented reduction without extra metadata.} To reduce the atomic operations at each lane end, MERBIT employs a warp-level segmented reduction that propagates partial sums together with dynamically generated row termination flags (Algorithm~\ref{warpsegmentedsum}). Unlike the segmented reduction in CSR5 (see Appendix~A, Algorithm~2), our approach requires no auxiliary $\textit{segment\_offset}$ array, preserving compact descriptors and reducing per-tile overhead.

\textit{(3) Dual-buffer commit to avoid full-vector reset.} Iterative SpMV requires $\bm{y}=\bm{0}$ at the start of each iteration, due to the utilization of atomic additions. Explicitly resetting $\bm{y}$ in global memory incurs nontrivial overhead. MERBIT introduces an auxiliary buffer $\bm{z}$ with the same size as $\bm{y}$ and alternates the roles of $\bm{y}$ and $\bm{z}$ across iterations, while only resetting block boundary locations during the commit phase (see Appendix~C, Algorithm~6). This can reduce global memory traffic.

\subsection{Parameter Selection}
MERBIT involves two parameters, $\omega$ and $\sigma$. The parameter $\omega$ is fixed by the hardware warp size (e.g., $\omega=32$ on NVIDIA GPU), whereas $\sigma$ is a tunable parameter that affects the performance of MERBIT in terms of the memory footprint of \textit{TILE}, the shared memory consumption per block and the fraction of fast-tackled tiles.

According to the earlier discussion, the shared memory requirement per thread block is determined by the length of merge-path segment assigned to this block. Denoted by $smem_{\sigma}$ the shared memory allocation per block, it holds that
\[
	smem_{\sigma} = (b + 1) \times \sigma \times sizeof(value), 
\]
where \textit{b} is the block size. Since each streaming multiprocessor (SM) of GPU has a fixed shared memory budget, larger $\sigma$ increases $smem_{\sigma}$ and may reduce the number of resident blocks per SM, potentially lowering occupancy and therefore hurting the performance.

Fast-tackled tile accelerates the computation. Let $r_f$ denote the fraction of tiles that are fast-tackled (i.e., spanning a single row). With $\sigma$ increasing, a tile is more likely to span multiple rows, which decreases $r_f$ and reduces the benefit of bypassing shared memory loading and lane descriptor parsing, and therefore degrading the performance.

Let $mem_{\sigma}$ denote the memory footprint of \textit{TILE}. Since both $\textit{tile}\_\textit{x}$ and $\textit{tile}\_\textit{y}$ contain $\lceil (m+n)/(\omega\sigma)\rceil+1$ 32-bit entries, $\textit{lane}\_\textit{desc}$ contains $\lceil (m+n)/\sigma\rceil$ 32-bit descriptors, we have \\
\[
	mem_{\sigma} = 8\Big(\Big\lceil \frac{m+n}{\omega\sigma}\Big\rceil + 1\Big) + 4\Big\lceil \frac{m+n}{\sigma}\Big\rceil(1 - r_{f}).
\]
Therefore, increasing $\sigma$ reduces the \textit{TILE}'s memory footprint, which is beneficial for memory-system-limited SpMV.

The above factors induce a three-way trade-off: larger $\sigma$ reduces \textit{TILE}'s memory footprint, but increases shared memory pressure and typically decreases $r_f$. Since tuning $\sigma$ for each matrix remains challenging, we choose a fixed $\sigma$ for each precision by prioritizing stable occupancy while keeping the scheduling metadata compact:
\begin{itemize}
	\item \textbf{Single precision:} shared memory is relatively sufficient for smaller data size; we set $\sigma=14$ to minimize the scheduling metadata traffic.
	\item \textbf{Double precision:} larger data size increases shared memory pressure; we set $\sigma=7$ to approximately match the occupancy of single-precision case.
\end{itemize}

\section{Experiments}
In this section, we experimentally evaluate MERBIT. We first describe the experimental setup, including the hardware configuration, software environment, benchmark suite and competitive baselines. We then report single SpMV performance in terms of computational throughput and speedup, followed by kernel profiling. Finally, we apply the kernels to iterative workloads to evaluate the end-to-end performance.

\subsection{Experimental setting}
All kernels are implemented in CUDA C++ using CUDA Toolkit V12.4 and compiled with the \texttt{-O3} optimization flag. Experiments are conducted on a server equipped with an NVIDIA GeForce RTX 4090 GPU running Ubuntu 20.04.1 LTS. All reported runtimes and speedups are measured under normal execution, without attaching NVIDIA Nsight Compute or any other profiler. NVIDIA Nsight Compute is used only in separate profiling runs. We collect 50 sparse matrices from the SuiteSparse Matrix Collection (see Appendix~D Table~I), with each one containing more than 10 million nonzeros. We compare MERBIT with eight competitive baselines, including vendor libraries (cuSPARSE CSR/COO), open-source libraries (Ginkgo CSR/COO/HYB), and academic approaches (Merge-Based SpMV, HOLA and CSR5). The source code of this work is available at \url{https://github.com/Zhangq388/MERBIT}.

\subsection{Single SpMV Performance}
\subsubsection{Computational Throughput}
\begin{figure}[h]
	\centering
	\includegraphics[width=0.90 \textwidth]{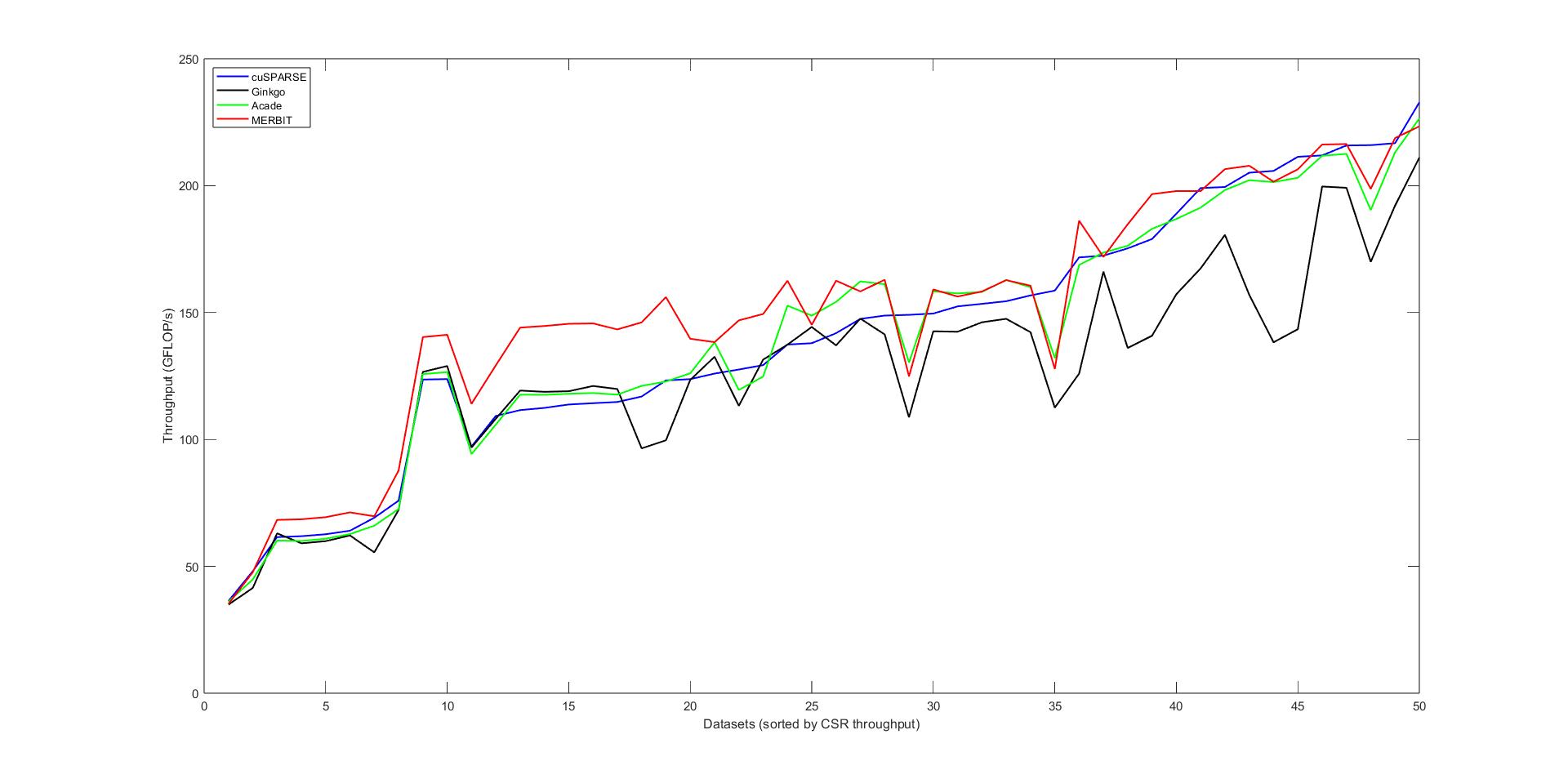}
	\caption{The throughput of SpMV kernels in single precision}
	\label{spmvsingle}
\end{figure}

\begin{figure}[h]
	\centering
	\includegraphics[width=0.90 \textwidth]{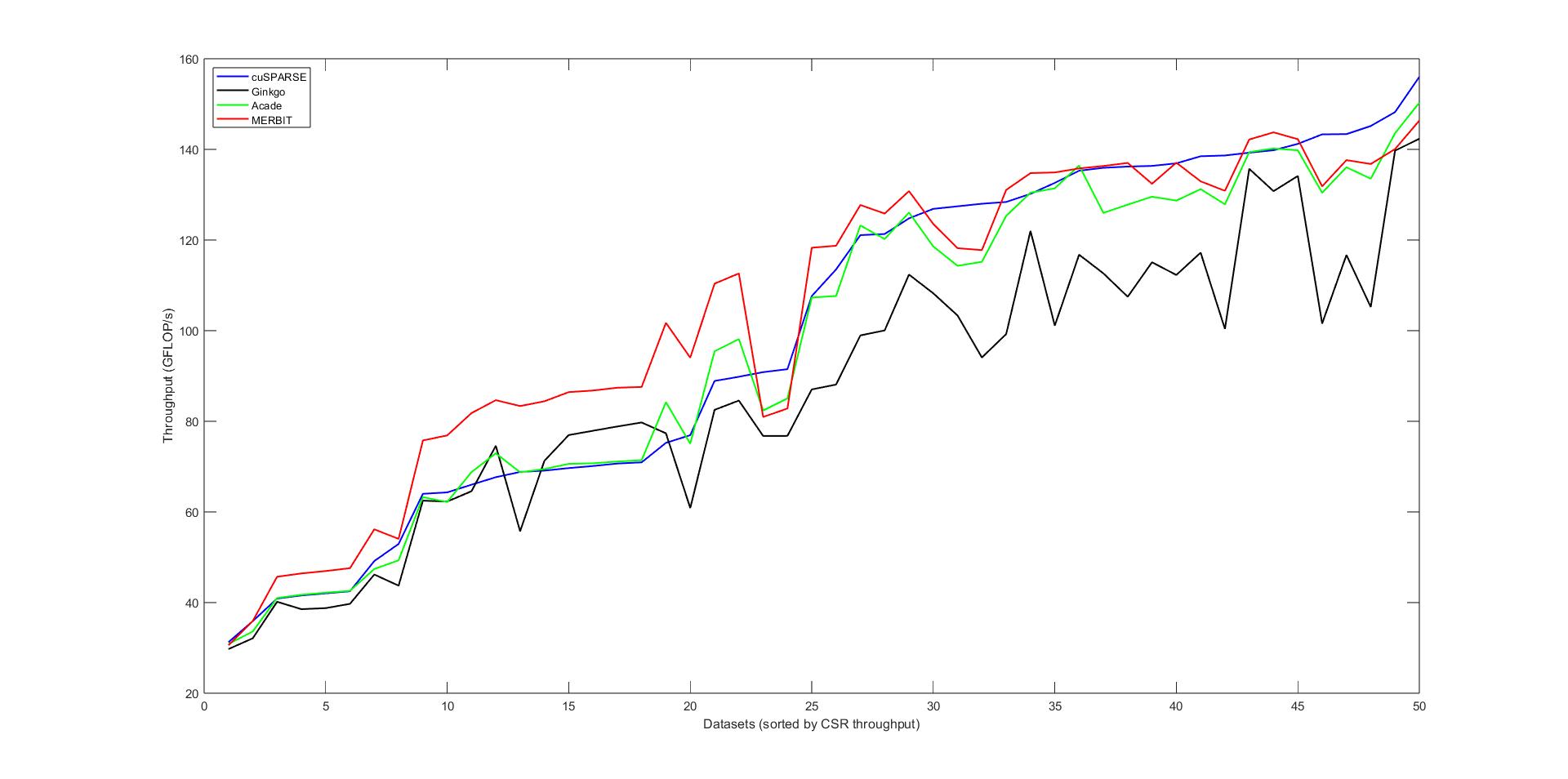}
	\caption{The throughput of SpMV kernels in double precision}
	\label{spmvdouble}
\end{figure}

We first evaluate the raw SpMV performance using computational throughput. For a sparse matrix with $m$ nonzero entries, the computational throughput is defined as
\[
	\textit{CT} = \frac{2m}{T},
\]
where $T$ is the average execution time of one SpMV call on GPU, and $m$ represents the number of nonzeros. In our experiments, $T$ is averaged over 400 SpMV iterations. Host--device data transfers are excluded, and the one-time preprocessing costs of CSR5 and MERBIT are not included in this single-SpMV measurement. Figure \ref{spmvsingle} and Figure \ref{spmvdouble} illustrate the per-dataset throughput in single and double precision, respectively. The datasets are sorted by the throughput of cuSPARSE CSR, which provides a consistent rank-based view from CSR-unfriendly cases to CSR-friendly cases. To reduce visual clutter, we also plot three aggregated baselines. \textit{cuSPARSE} denotes the per-dataset maximum throughput among cuSPARSE CSR and cuSPARSE COO; \textit{Ginkgo} denotes the per-dataset maximum throughput among Ginkgo CSR, COO, and HYB; and \textit{Acade} denotes the per-dataset maximum throughput among the academic methods, including Merge-Based SpMV, HOLA, and CSR5. Table~\ref{table2} further reports the number of datasets on which each individual kernel achieves the highest throughput. We summarize the results as below:
\begin{enumerate}
	\item[(1)] MERBIT delivers the best overall performance, achieving the highest throughput on 36 datasets in single precision and 35 datasets in double precision. These results highlight the benefit of merge-path tiling together with compact bit-field encoding. cuSPARSE is also competitive, leading on 9 (single) and 14 (double) datasets. In contrast, Ginkgo attains the lowest throughput on the majority of datasets in our evaluation. The gap between cuSPARSE and Ginkgo suggests that engineering-level optimizations are crucial: even when implementations target similar formats (e.g., CSR/COO-based SpMV), performance can differ substantially depending on low-level optimizations.
	\item[(2)] MERBIT is more effective on datasets where cuSPARSE CSR delivers relatively lower throughput. In contrast, when CSR attains higher throughput, MERBIT tends to be less competitive. This pattern suggests that merge-path tiling primarily pays off by reducing the irregularity-induced overhead, which becomes less beneficial on CSR-friendly datasets. Among academic approaches, HOLA is more effective in single precision than in double precision, this is consistent with its non-coalesced matrix access for long rows and higher synchronization overhead in implementation.
\end{enumerate}  
\begin{table}[H]
	\caption{The number of datasets where SpMV kernels achieve the best\label{table2}}
	\centering
	\resizebox{0.90 \textwidth}{!}{
		\begin{tabular}{c c c c c c c c c c}
			\toprule
			Precision  &COO  &CSR  &GCSR &GCOO  &GHYB  &MERGE  &HOLA  &CSR5 &MERBIT      \\
			\midrule
			single     &0    &9    &0    &0     &0     &0      &5     &0    &\textbf{36} \\
			double     &0    &14   &0    &0     &0     &0      &1     &0    &\textbf{35} \\ 
			\bottomrule
	\end{tabular}}
\end{table}

\subsubsection{Speedup}
We further evaluate the relative performance of different SpMV kernels using cuSPARSE COO as the baseline. For a kernel $K$ on a dataset $A_i$, the speedup is defined as
\[
	\textit{speedup}_{K}(A_i) = \frac{T_{\text{COO}}(A_i)}{T_{K}(A_i)},
\]
where $T_{\text{COO}}(A_i)$ and $T_K(A_i)$ denote the average execution time of cuSPARSE COO and kernel $K$, respectively. A speedup larger than 1 indicates that the corresponding kernel is faster than the
cuSPARSE COO baseline. Figure~8 and Figure~9 show the speedup distributions in single and double precision, respectively. Table~3 summarizes the corresponding descriptive statistics. We highlight the following observations:
\begin{enumerate}
	\item[(1)] MERBIT achieves the most stable and competitive performance among all evaluated kernels. In single precision, MERBIT achieves a minimum speedup of $1.00 \times$, a median speedup of $1.27 \times$, and a geometric mean speedup of $1.27 \times$. In double precision, MERBIT further achieves a minimum speedup of $1.02 \times$, a median speedup of $1.28 \times$, and a geometric mean speedup of $1.25 \times$. These results show that MERBIT not only improves average performance, but also avoids severe slowdowns across the benchmark suite.
	\item[(2)] cuSPARSE CSR achieves the largest maximum speedup in both precisions. This indicates that highly optimized row-based kernels can be very effective on matrices with relatively regular row structures. However, its minimum speedup is lower than that of MERBIT, especially in single precision, suggesting that CSR-based execution is more sensitive to irregular sparsity. In contrast, Merge-Based SpMV and MERBIT
	both keep their minimum speedups close to $1.00 \times$, which confirms the robustness of merge-path-based workload partitioning. Since merge-path partitioning balances both nonzero computations and row-boundary events, it can avoid the severe workload imbalance that often appears in row-based kernels.
\end{enumerate}

To better understand when MERBIT is most effective, we further stratify the 50 datasets by their average degree $d=m/n$ using thresholds tied to $\sigma$: in single precision, G-L ($d \le 14$) and G-H ($d > 14$) contains 35 and 15 datasets, respectively; in double precision, G-L ($d \le 7$) and G-H ($d > 7$) contains 24 and 26 datasets, respectively. Table \ref{grouped_speedup} reports (i) the number of datasets where each kernel outperforms MERBIT (\textit{outperf}); (ii) the geometric mean speedup (\textit{gmean}). We highlight the following observations:
\begin{enumerate}
	\item[(1)] MERBIT is particularly effective on low-degree matrices. In the G-L group, MERBIT achieves a geometric mean speedup of $1.25 \times$ in both single and double precision, and almost no competing kernel outperforms it. This is because low-degree graph matrices usually contain many short rows and irregular row boundaries. In this case, row-based execution suffers from insufficient work per row and load imbalance, whereas MERBIT partitions the work in the two-dimensional space of rows and nonzeros. As a result, it balances computation more effectively while keeping \textit{TILE} compact.
	\item[(2)] In the high-degree group, the performance gap becomes smaller. For example, cuSPARSE CSR achieves a geometric mean speedup of $1.32 \times$ in single precision and $1.27 \times$ in double precision on G-H, and it outperforms MERBIT on several datasets. This is expected because high-degree matrices usually provide more work per row, making row-based execution more efficient and reducing the relative benefit of merge-path partitioning. Meanwhile, the metadata overhead of \textit{TILE} becomes more noticeable when the average degree is high, since the compact merge-path schedule may no longer be much lighter than the CSR row-offset structure.
\end{enumerate}
\begin{figure}[H]
	\centering
	\includegraphics[width=0.95 \textwidth]{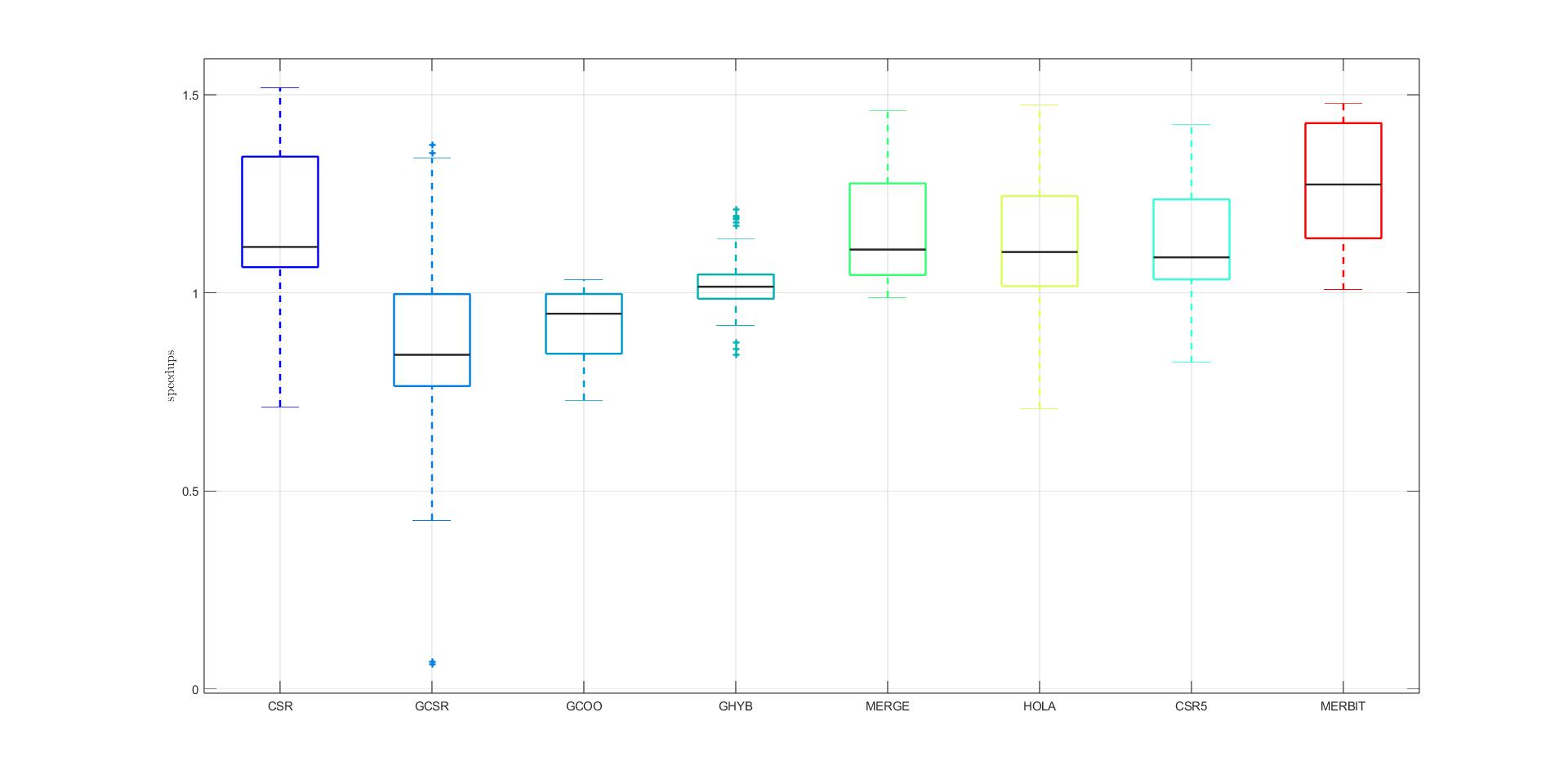}
	\caption{Speedup of SpMV kernels in single precision}
	\label{speedupsingle}
\end{figure}
\begin{figure}[H]
	\centering
	\includegraphics[width=0.95 \textwidth]{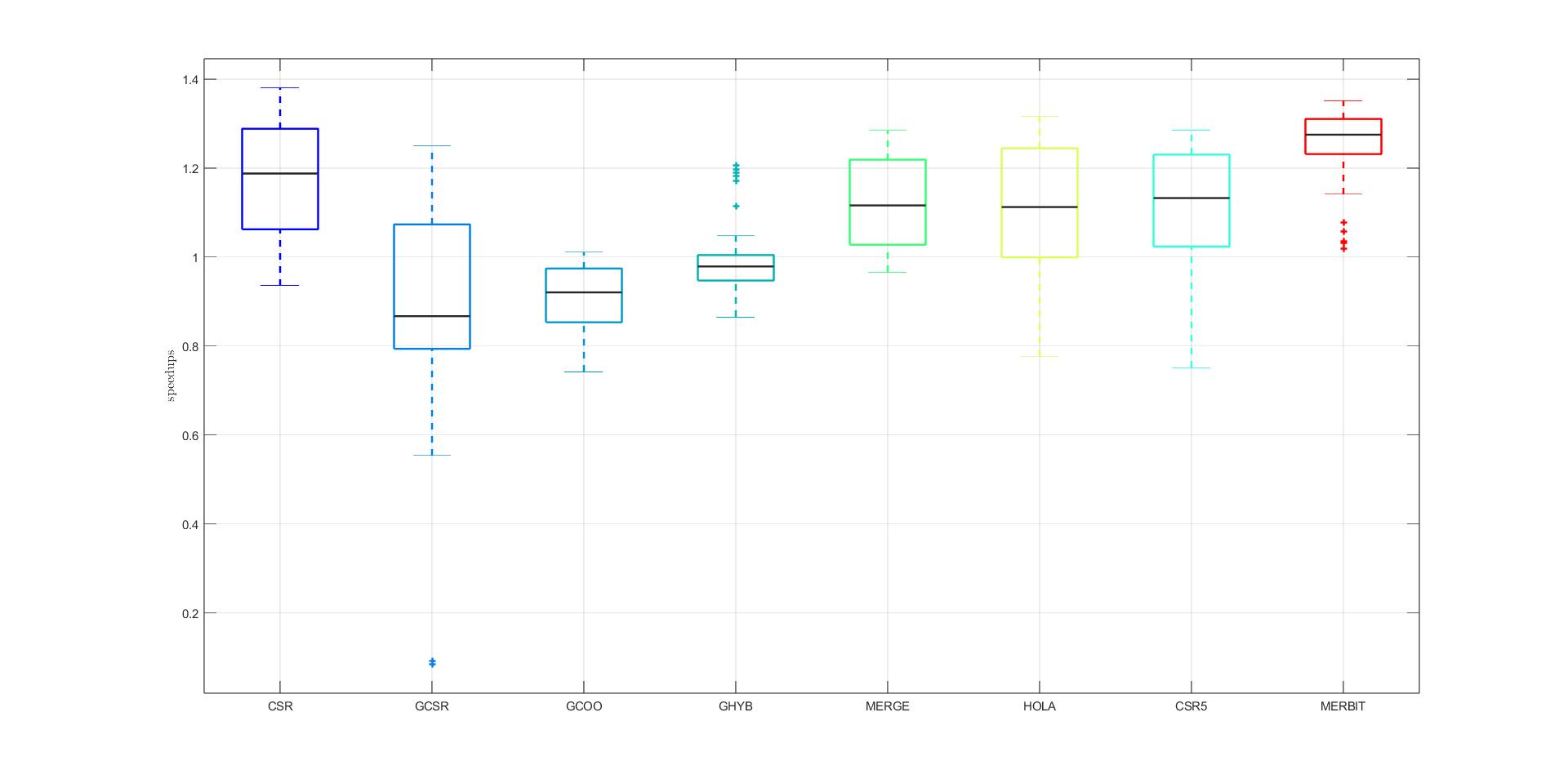}
	\caption{Speedup of SpMV kernels in double precision}
	\label{speedupdouble}
\end{figure}
\begin{table}[H]
	\caption{The descriptive statistics on speedup of SpMV kernels\label{table3}}
	\centering
	\resizebox{0.80 \textwidth}{!}{
		\begin{tabular}{c c c c c c c c c}
			\toprule
			\multirow{2}*{SpMV}                   &\multicolumn{4}{c}{single}                                    &\multicolumn{4}{c}{double}    \\
			\cmidrule(lr){2-9}
			&$min$         &$max$         &$median$      &$gmean$        &$min$         &$max$           &$median$      &$gmean$         \\
			\midrule
			CSR     &0.71          &\textbf{1.52} &1.12          &1.16           &0.94          &\textbf{1.38}   &1.19          &1.17           \\
			GCSR    &0.06          &1.37          &0.84          &0.79           &0.08          &1.25            &0.87          &0.83           \\
			GCOO    &0.73          &1.03          &0.95          &0.91           &0.74          &1.01            &0.92          &0.90           \\
			GHYB    &0.84          &1.21          &1.02          &1.02           &0.86          &1.21            &0.98          &0.99           \\
			MERGE   &0.99          &1.46          &1.11          &1.15           &0.97          &1.28            &1.12          &1.12           \\
			HOLA    &0.71          &1.47          &1.10          &1.12           &0.78          &1.32            &1.11          &1.10           \\
			CSR5    &0.82          &1.43          &1.09          &1.12           &0.75          &1.29            &1.13          &1.10           \\
			MERBIT  &\textbf{1.00} &1.48          &\textbf{1.27} &\textbf{1.27}  &\textbf{1.02} &1.35            &\textbf{1.28} &\textbf{1.25}  \\
			\bottomrule
	\end{tabular}}
\end{table}
\begin{table}[H]
	\caption{Speedup grouped by average degree $d=m/n$\label{grouped_speedup}}
	\centering
	\footnotesize
	\resizebox{0.90 \textwidth}{!}{
		\begin{tabular}{c|cc|cc|cc|cc}
			\toprule
			\multirow{3}{*}{Kernel}       &\multicolumn{4}{c|}{single} &\multicolumn{4}{c}{double}                             \\
			&\multicolumn{2}{c|}{G-L (35)} &\multicolumn{2}{c|}{G-H (15)}   &\multicolumn{2}{c|}{G-L (24)}  &\multicolumn{2}{c}{G-H (26)}      \\
			&\textit{outperf}   &$gmean$          &\textit{outperf}     &$gmean$          &\textit{outperf}   &$gmean$       &\textit{outperf}  &$gmean$  \\
			\midrule
			CSR    &4                  &1.10             &6                    &\textbf{1.32}    &1                  &1.07          &13                &\textbf{1.27}  \\
			GCSR   &0                  &0.68             &0                    &1.13             &0                  &0.66          &1                 &1.03           \\
			GCOO   &0                  &0.90             &0                    &0.96             &0                  &0.84          &0                 &0.96           \\
			GHYB   &0                  &1.04             &0                    &0.98             &0                  &1.01          &0                 &0.97           \\
			MERGE  &0                  &1.10             &1                    &1.26             &0                  &1.06          &1                 &1.19           \\
			HOLA   &3                  &1.05             &6                    &1.28             &0                  &0.98          &6                 &1.21           \\
			CSR5   &2                  &1.07             &0                    &1.23             &0                  &1.01          &3                 &1.20           \\
			MERBIT &-                  &\textbf{1.25}    &-                    &1.30             &-                  &\textbf{1.25} &-                 &1.25           \\
			\bottomrule
	\end{tabular}}
\end{table}
\subsection{Kernel profiling}
This subsection uses profiling metrics to explain the performance trends observed in Section~5.2. To better understand the performance differences, we profile the primary SpMV kernels on all 50 datasets in our benchmark suite, in both single and double precision. We exclude Ginkgo HYB because its SpMV execution is split into two kernels, which prevents a direct kernel comparison. Table~\ref{profile} reports metrics averaged across the 50 datasets and groups them into three categories: 
\begin{enumerate}
	\item [(1)] \textbf{Memory behavior}, including the achieved memory throughput (\emph{Thr}), memory busy utilization (\emph{Busy}), memory pipelines utilization (\emph{Pipes}), coalescing rate (\emph{Coal}), and normalized DRAM read bytes (\textit{DRAM}). \emph{Coal} is defined as the ratio of L1 load requests to L1 load sectors (requests/sectors). A higher value indicates that fewer sectors are required per request on average, i.e., the memory accesses are more coalesced. The DRAM read bytes are normalized to cuSPARSE COO.
	\item [(2)] \textbf{Warp state}, including the \emph{Active} and \emph{Eligible} warps per scheduler, executed instructions per warp cycle (\emph{IPC}), and warp cycles per instruction (\emph{WCPI}). These metrics reflect whether resident warps can continuously issue instructions or are frequently blocked by memory dependencies and synchronization.
	\item [(3)] \textbf{Warp stall}, reporting the dominant stall reasons, including long scoreboard stalls (\textit{Long}), short scoreboard stalls (\textit{Short}), memory pipe throttle stalls (\textit{MIO}), and barrier stalls (\textit{Bar}). These metrics help identify whether the performance loss mainly comes from memory latency, memory-pipeline pressure, or synchronization overhead.
\end{enumerate}
\subsubsection{Memory Behavior}
All kernels show clear memory-system pressure, as \emph{Throughput} (except GCOO) stays within a narrow range (83\%--89\%). GCOO achieves the lowest \emph{Throughput} and \emph{Busy}, while exhibiting the highest \emph{Pipes}, \emph{Coalesce}, and \emph{DRAM}. This behavior is expected because GCOO additionally loads a consecutive \textit{row}\_\textit{ind} array, increasing memory traffic and pressure on the memory pipelines. CSR5 shows substantially lower \emph{Pipes} (9.95\% and 7.63\%), whereas its other memory metrics remain close to the average level. A plausible explanation is its iterative load--compute flow introduces more long-latency dependencies; as a result, warps spend more cycles waiting and the overlap between memory operations and computation is reduced. HOLA attains the lowest \emph{Coalesce} (7.82\% and 7.80\%), yet its \emph{DRAM} remains close to the average level. This is consistent with its long-row handling, which is less coalesced but does not necessarily translate into higher DRAM reads due to its overall access pattern. MERBIT achieves higher \emph{Coalesce} (14.43\% and 13.44\%) and the lowest \emph{DRAM} (73\% and 82\%), while keeping the remaining metrics close to the average. Staging products in shared memory allows MERBIT to load the matrix stream with more contiguous global memory access, thereby reducing the number of sector transactions per request and lowering DRAM traffic. 

\subsubsection{Warp State}
Across all kernels, \emph{Eligible} remains below 3\% of \emph{Active} per scheduler, indicating that although many warps are resident, only a small fraction are ready to issue, i.e., most warps frequently stall. CSR5 exhibits the lowest \emph{Eligible} (0.09 and 0.07) together with a higher \emph{WCPI}, which aligns with its memory behavior: warps and instructions often wait on long-latency data dependencies, reducing issue readiness. HOLA exhibits the largest cross-precision variation in \emph{WCPI}, increasing from 58.09 (single) to 107.82 (double), which we attribute to frequent synchronization required by its multiplication and warp-level transpose. MERBIT achieves highest \emph{IPC} (excluding GCOO) (0.53 and 0.55) and a lower \emph{WCPI} (57.93 and 61.69), suggesting that merge-path tiling, shared-memory staging, and buffered writes improve pipeline utilization by providing more continuous instruction issue opportunities. 

\subsubsection{Warp Stall}
Long-scoreboard stalls dominate across all kernels, confirming that SpMV performance is primarily constrained by memory dependency and latency. CSR5 exhibits the highest \emph{Long} component (88\% and 86\%), consistent with its low \emph{Pipes}, \emph{Eligible}, and \emph{IPC}, as well as its higher \emph{WCPI}. In contrast, HOLA shows a pronounced barrier-stall component (29\% and 26\%), which aligns with its elevated \emph{WCPI} and indicates frequent synchronization. MERBIT presents a more balanced stall breakdown, avoiding extreme stall dominance and thereby sustaining lower \emph{WCPI}.

Overall, high GPU SpMV performance is largely determined by reducing the dominant inefficiencies, rather than optimizing a single isolated metric. 

\subsubsection{Implications on modern GPUs}
Modern GPUs continue to evolve with larger L1/L2 caches, higher memory bandwidth, larger register files, higher shared-memory capacity, and asynchronous global-to-shared transfers such as \texttt{cp.async}. MERBIT is expected to remain effective on future architectures. This expectation is supported by our profiling results, which show that MERBIT reduces irregular-access overheads, rather than relying on a single architecture-specific feature. In particular, a larger L2 cache can improve the L2 hit rate, higher bandwidth can increase effective data-movement efficiency, and larger SM resources (registers and shared memory) can raise achievable occupancy.

Moreover, increases in register-file size and shared-memory capacity can expand the feasible design space of MERBIT’s tunable parameter $\sigma$. As shown in Table~\ref{profile}, MERBIT exhibits relatively lower \emph{Active} (7.83 and 7.81) because the number of resident thread blocks per SM is constrained by register and shared-memory usage. With more registers and shared memory per SM, larger $\sigma$ values and 64-bit \textit{lane}\_\textit{desc} become feasible, which may extend MERBIT’s advantage to matrices with higher degree.

Finally, we expect the benefit of \texttt{cp.async} to be limited for MERBIT in our setting. First, MERBIT’s critical bottleneck is typically the irregular gather from $x$, whereas \texttt{cp.async} is most effective for pipelining contiguous global-to-shared transfers. Second, MERBIT consumes staged intermediate products quickly, leaving limited opportunity to overlap enough memory latency with computation. This is consistent with our preliminary experiments: applying \texttt{cp.async} to stage values/indices yields little to no speedup.
\begin{table*}[!t]
	\caption{Kernel profiling summary averaged over the benchmark suite.\label{profile}}
	\centering
	\footnotesize
	\resizebox{0.98 \textwidth}{!}{
		\begin{tabular}{c c c c c c c | c c c c | c c c c}
			\toprule
			\multirow{2}{*}{Precision} & \multirow{2}{*}{Kernel} & \multicolumn{5}{c}{Memory (\%)} & \multicolumn{4}{c}{Warp state} & \multicolumn{4}{c}{Warp stall (\%)} \\
			\cmidrule(lr){3-7}\cmidrule(lr){8-11}\cmidrule(lr){12-15}
			&           &Thr           &Busy      &Pipes         &Coal          &DRAM          &Active  &Eligible  &IPC      &WCPI     &Long  &Short  &MIO   &Bar     \\
			\midrule
			\multirow{6}{*}{single}
			&CSR        &84.41         &58.05     &15.67         &14.50         &81            &11.19   &0.15      &0.37     &132.57   &48    &7      &17    &18     \\
			&COO        &74.48         &45.96     &13.16         &14.50         &100           &11.02   &0.16      &0.47     &127.28   &73    &9      &11    &0      \\
			&GCSR       &88.82         &48.73     &10.96         &9.50          &82            &10.52   &0.12      &0.23     &176.91   &29    &8      &22    &16     \\
			&GCOO       &67.32         &36.95     &20.57         &27.89         &110           &10.54   &0.35      &0.89     &40.98    &79    &3      &0     &0      \\
			&MERGE      &88.04         &60.67     &19.71         &12.32         &77            &11.78   &0.23      &0.50     &104.92   &37    &11     &23    &21     \\
			&HOLA       &86.05         &53.71     &16.38         &7.82          &77            &5.87    &0.15      &0.44     &58.09    &38    &8      &12    &29     \\
			&CSR5       &85.94         &58.55     &9.95          &11.68         &75            &10.06   &0.09      &0.31     &148.37   &88    &3      &2     &0      \\
			&MERBIT     &87.03         &59.97     &15.22         &14.43         &73            &7.83    &0.19      &0.53     &66.37    &50    &11     &12    &14     \\
			\midrule
			\multirow{6}{*}{double}
			&CSR        &81.01         &47.81     &12.18         &11.28         &89            &8.29    &0.11      &0.29     &122.61   &28    &18     &25    &17     \\
			&COO        &74.52         &41.63     &12.97         &13.84         &100           &11.11   &0.12      &0.39     &152.85   &70    &12     &13    &0     \\
			&GCSR       &89.08         &48.89     &13.39         &8.93          &91            &9.59    &0.13      &0.29     &144.84   &29    &15     &24    &18     \\
			&GCOO       &76.36         &44.46     &20.51         &23.64         &114           &10.54   &0.31      &0.85     &53.44    &80    &7      &0     &0      \\
			&MERGE      &85.59         &53.77     &17.68         &10.37         &83            &11.48   &0.21      &0.44     &113.12   &22    &19     &34    &16     \\
			&HOLA       &83.76         &49.94     &13.75         &7.80          &83            &9.75    &0.18      &0.40     &107.82   &20    &19     &27    &26     \\
			&CSR5       &85.43         &56.09     &7.63          &9.36          &82            &10.05   &0.07      &0.24     &193.46   &86    &4      &3     &0     \\
			&MERBIT     &85.98         &52.50     &15.92         &13.44         &82            &7.81    &0.18      &0.55     &61.80    &54    &18     &4     &12     \\
			\bottomrule
	\end{tabular}}
\end{table*}
\subsection{Parameter Optimization}
\begin{figure}[!t]
	\centering
	\includegraphics[width=0.90 \textwidth]{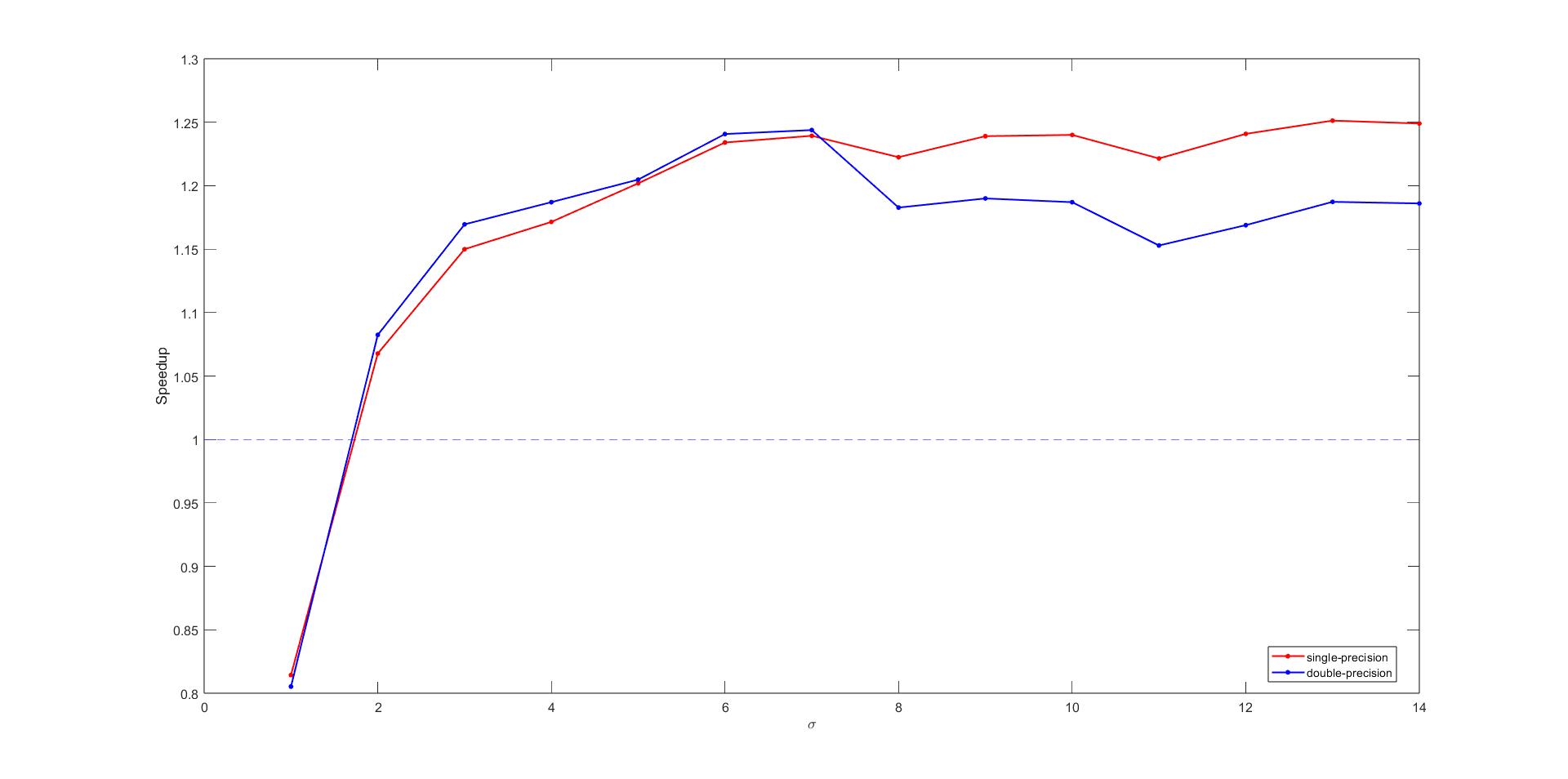}
	\caption{Speedup of MERBIT with different $\sigma$ value}
	\label{parameteroptimization}
\end{figure}

We validate the effectiveness of the proposed parameter optimization strategy for $\sigma$. Figure~\ref{parameteroptimization} reports the average speedup of MERBIT normalized to cuSPARSE COO under different $\sigma$ values:  
\begin{enumerate}
	\item [(1)] In single precision, the speedup increases with $\sigma$ growing. At $\sigma=1$, MERBIT employs the finest-grained tiles, which incurs substantial scheduling and boundary-handling overhead, leading to a speedup below $1$. The best average performance is achieved at $\sigma=13$. This trend is consistent with our analysis: a larger $\sigma$ reduces the footprint of \textit{TILE} and amortizes per-tile overhead, while the shared memory pressure remains affordable.
	\item [(2)] In double precision, the speedup increases up to $\sigma=7$ and then slightly decreases. The larger element size increases the shared memory requirement and reduces achieved occupancy, which limits latency hiding when $\sigma$ becomes large. This trend is also consistent with our profiling results: when $\sigma=14$ in double precision, the average achieved occupancy per SM drops noticeably (23.50) compared to $\sigma=7$ (30.15), explaining the degradation beyond $\sigma=7$.
\end{enumerate}
Although $\sigma=13$ yields the best average in single precision, its improvement over $\sigma=14$ is marginal. Overall, $\sigma=14$ (single) and $\sigma=7$ (double) offer robust default settings.

\subsection{Performance of applying to Iterative Workloads}
To evaluate the practical applicability beyond single SpMV, we apply each kernel (except Ginkgo HYB due to insufficient device memory) to two representative iterative workloads: PageRank and the Bi-Conjugate Gradient Stabilized (BiCGSTAB)~\cite{van1992bi}. For both workloads, we report the end-to-end runtime
\[
	T = T_{p} + T_{r},
\]
where $T_{p}$ is the one-time preprocessing cost, and $T_{r}$ is the runtime of $r$ iterations. 
\subsubsection{PageRank}
We compute PageRank using the Power method with damping factor $c=0.85$. Let $\boldsymbol{\pi}(r)$ be the PageRank vector of the $r$-th iteration. Let $\boldsymbol{\pi}^{\star}$ be a high-accuracy reference solution obtained by running the power method for 210 iterations. We measure the relative $\ell_\infty$ error as
\[
	ERR = \lVert \bm{r} \rVert_\infty,\quad r_i=\frac{\pi^{(r)}_i-\pi^{\star}_i}{\pi^{\star}_i},\ i=1,2,\ldots,n.
\]
Table \ref{table5} reports $T$ on eight representative datasets under the convergence criterion $ERR < 10^{-10}$, where the value in parentheses denotes $T_p$. Table \ref{table6} reports the geometric mean speedups normalized to cuSPARSE COO under different iteration counts, the value in parentheses denotes the number of datasets (out of 50) where the corresponding kernel outperforms MERBIT. The results are summarized as below: 
\begin{enumerate}
	\item [(1)] MERBIT incurs substantially lower preprocessing overhead than CSR5. Averaged over all 50 datasets, CSR5 preprocessing costs up to $3.17\times$ the time of a single SpMV, whereas MERBIT requires only $1.29\times$. As the iteration counts grow, the averaged speedups of CSR5 and MERBIT increase, benefiting from amortization. 
	\item [(2)] MERBIT consistently outperforms the Ginkgo library and prior academic approaches, and is competitive with cuSPARSE CSR. Compared to the single SpMV, the observed speedups of most methods decrease in PageRank ($1.25\times$ to $1.18\times$ for MERBIT), indicating additional iteration-level overhead such as vector-wide updates.
\end{enumerate}
\begin{table}[!t]
	\caption{The end-to-end runtime (.ms) of PageRank\label{table5}}
	\footnotesize
	\centering
	\resizebox{0.90 \textwidth}{!}{
		\begin{tabular}{l c c c c c c c c}
			\toprule
			Dataset            &COO    &CSR            &GCSR   &GCOO    &MERGE    &HOLA    &CSR5                &MERBIT                \\
			\midrule
			italy\_osm         &84.1   &79.5           &97.8   &97.5    &79.5     &84.7    &80.6(1.6)           &\textbf{70.7(0.3)}    \\
			road\_usa          &315    &294            &358    &364     &301      &311     &310.2(5.2)          &\textbf{261.5(2.5)}   \\
			soc-LiveJournal1   &250    &205            &229    &281     &226      &220     &227.8(4.8)          &\textbf{213.5(2.5)}   \\
			kmer\_V2a          &1070   &1060           &1400   &1350    &1030     &1160    &1329.3(9.3)         &\textbf{958.0(5.0)}   \\
			mawi\_201512020130 &1830   &1740           &2110   &2080    &1830     &1840    &1763.2(13.2)        &\textbf{1525.2(5.2)}  \\
			uk-2002            &1050   &851            &903    &1090     &855     &867     &874(15)             &\textbf{836.4(5.4)}     \\
			soc-sinaweibo      &2110   &\textbf{1810}  &2150   &2490    &2090     &2010    &2135.8(15.8)        &2054.7(4.7)             \\
			arabic-2005        &2090   &1650           &1690   &2150    &1660     &1660    &1679.6(19.6)        &\textbf{1625.2(5.2)}    \\
			\bottomrule
	\end{tabular}}
\end{table}
\begin{table}[!t]
	\caption{Average PageRank speedups under different iteration counts.\label{table6} }
	\footnotesize
	\centering
	\resizebox{0.90 \textwidth}{!}{
		\begin{tabular}{c c c c c c c c}
			\toprule
			iters &CSR          &GCSR       &GCOO      &MERGE       &HOLA        &CSR5         &MERBIT          \\
			\midrule
			50    &1.170(26)    &0.869(2)   &0.876(0)  &1.103(11)   &1.068(14)   &1.006(2)     &\textbf{1.158}  \\ 
			100   &1.169(25)    &0.886(2)   &0.892(0)  &1.103(4)    &1.067(8)    &1.037(2)     &\textbf{1.174}  \\ 
			150   &1.169(25)    &0.891(2)   &0.899(0)  &1.103(1)    &1.067(7)    &1.048(2)     &\textbf{1.179}  \\ 
			200   &1.168(24)    &0.893(2)   &0.901(0)  &1.103(1)    &1.066(6)    &1.053(2)     &\textbf{1.182}  \\ 
			\bottomrule
	\end{tabular}}
\end{table}

\subsubsection{Bi-Conjugate Gradient Stabilized}
BiCGSTAB is employed to solve the linear system $\bm{A}\bm{x}=\bm{b}$. 
In our evaluation, we construct $\bm{A}$ from the undirected graphs by setting all its nonzero entries to 1. 
To generate a consistent $\bm{b}$, we sample a random $\bm{x}$ with entries uniformly drawn from $[-1,1]$, and then compute $\bm{b}=\bm{A}\bm{x}$. 
We measure the relative residual as
\[
	tol = \frac{\lVert \bm{A}\bm{x}^{(r)} - \bm{b} \rVert_2}{\lVert \bm{b} \rVert_2},
\]
and terminate when $tol < 10^{-10}$ or the iterations reach 20000. Due to insufficient device memory, BiCGSTAB successfully executes on only 21 matrices in our evaluation (see Appendix~D Table~I). Table~\ref{cgstable1} reports the end-to-end runtime $T$ on four representative datasets, and Table~\ref{cgstable2} reports the geometric mean speedups normalized to cuSPARSE COO, the value in parentheses denotes the number of datasets (out of 21) where the corresponding kernel outperforms MERBIT. The results are summarized as follows:
\begin{enumerate}
	\item [(1)] MERBIT achieves the best end-to-end performance in BiCGSTAB, consistently improving over cuSPARSE COO and other baselines. 
	\item [(2)] The overall speedup trend in BiCGSTAB is smaller than in PageRank, consistent with that BiCGSTAB iterations include multiple vector operations in addition to SpMV, which limits the attainable speedup from accelerating SpMV alone. 
\end{enumerate}
\begin{table}[H]
	\caption{The end-to-end runtime (.s) of BiCGSTAB\label{cgstable1}}
	\footnotesize
	\centering
	\resizebox{0.90 \textwidth}{!}{
		\begin{tabular}{l c c c c c c c c}
			\toprule
			Dataset            &COO       &CSR         &GCSR      &GCOO       &MERGE    &HOLA      &CSR5       &MERBIT            \\
			\midrule
			italy\_osm         &51.67     &50.21       &54.63     &54.64      &49.76    &51.41     &49.71      &\textbf{46.97}    \\
			road\_usa          &202.95    &196.97      &212.66    &214.35     &198.04   &200.75    &199.61     &\textbf{186.02}   \\
			kmer\_V2a          &537.76    &532.03      &623.79    &610.29     &529.53   &565.59    &608.57     &\textbf{508.59}   \\
			mawi\_201512020130 &1054.34   &1031.87     &1108.75   &1101.56    &1049.87  &1054.01   &1035.06    &\textbf{973.23}  \\
			\bottomrule
	\end{tabular}}
\end{table}
\begin{table}[H]
	\caption{Average BiCGSTAB speedups\label{cgstable2} }
	\footnotesize
	\centering
	\resizebox{0.90 \textwidth}{!}{
		\begin{tabular}{c c c c c c c}
			\toprule
			CSR           &GCSR       &GCOO      &MERGE      &HOLA        &CSR5         &MERBIT          \\
			\midrule
			1.099(8)      &0.988(2)   &0.931(0)  &1.062(1)   &1.048(2)    &1.047(2)     &\textbf{1.105}  \\ 
			\bottomrule
	\end{tabular}}
\end{table}
\section{Conclusion}
While accelerating iterative workloads on irregular matrices remains challenging, decreasing the time cost of single SpMV is a feasible approach. In this paper, we propose MERBIT---an innovative GPU SpMV initially motivated by PageRank but applicable to other iterative workloads. MERBIT can be viewed as an enhanced version of CSR5, where the executing flow and parallel scheme are implied in the merge-path, or as an extension of Merge-Based SpMV with a preprocessing which encodes merge-path into a compact bit-field representation. Additionally, three optimization strategies are employed. The experimental results show that MERBIT provides consistent benefits for repeated SpMV on irregular, graph-like matrices, especially when the average degree is low to moderate and the preprocessing cost can be amortized across iterations. However, for large datasets, the performance bottleneck primarily lies in memory access, including loading matrix and input vector. Future work will focus on improving the locality of input-vector accesses, which remains the major performance bottleneck for irregular SpMV.

\section*{Acknowledgments}
The authors are grateful to the anonymous reviewers for their valuable and helpful comments on improving the manuscript.

\bibliographystyle{elsarticle-num} 
\bibliography{ref}

\end{document}